\documentclass[a4paper, fleqn, usenatbib, useAMS]{mnras}

\usepackage{newtxtext, newtxmath}
\usepackage[T1]{fontenc}
\usepackage{ae, aecompl}

\usepackage{graphicx}
\DeclareGraphicsExtensions{.pdf, .ps, .eps}
\usepackage[multidot]{grffile}

\usepackage{xspace}
\usepackage[usenames, dvipsnames]{xcolor}

\newcommand{\starlight}{\textsc{starlight}\xspace}
\newcommand{\hii}{H\thinspace\textsc{ii}\xspace}
\newcommand{\Ha}{\ifmmode \mathrm{H}\alpha \else H$\alpha$\fi\xspace}
\newcommand{\Hb}{\ifmmode \mathrm{H}\beta \else H$\beta$\fi\xspace}
\newcommand{\nii}{\ifmmode [\mathrm{N}\,\textsc{ii}] \else [N~{\scshape ii}]\fi\xspace}
\newcommand{\oiii}{\ifmmode [\rm{O}\,\textsc{iii}] \else [O\,{\sc iii}]\fi\xspace}
\newcommand{\sii}{\ifmmode [\mathrm{S}\,\textsc{ii}] \else [S~{\scshape ii}]\fi\xspace}
\newcommand{\oi}{\ifmmode [\rm{O}\,\textsc{i}] \else [O\,{\sc i}]\fi\xspace}
\newcommand{\oii}{\ifmmode [\rm{O}\,\textsc{ii}] \else [O\,{\sc ii}]\fi\xspace}
\newcommand{\nOne}{\ifmmode [\mathrm{N}\,\textsc{i}] \else [N~{\scshape i}]\fi\xspace}
\newcommand{\siii}{\ifmmode [\mathrm{S}\,\textsc{iii}] \else [S~{\scshape iii}]\fi\xspace}

\title[Supernova remnants in NGC 4030]
      {Detection of supernova remnants in NGC 4030}

\author[Cid Fernandes et.\ al.]
       {R. Cid Fernandes$^1$,
	M. S. Carvalho$^1$,
	S. F. S\'anchez$^2$,
	A. de Amorim$^1$, 
	D. Ruschel-Dutra$^1$ \\
         $^1$Departamento de F\'{\i}sica - CFM - Universidade Federal de Santa Catarina,
        C.P. 476, 88040-900,  Florian\'opolis, SC, Brazil\\
	 $^2$Instituto de Astronom\'{\i}a, Universidad Nacional Aut\'onoma de M\'exico, A. P. 70-264, C.P. 04510 M\'exico, D.F., M\'exico
       }

\begin{document}

\maketitle

\begin{abstract}
MUSE-based emission-line maps of the spiral galaxy NGC 4030 reveal the existence of  unresolved sources with forbidden line emission enhanced with respect to those seen in its own \hii regions.  This study reports our efforts to detect and isolate  these objects and identify their nature. Candidates are first detected as unresolved sources on an image of the second principal component of the \Hb, \oiii5007, \Ha, \nii6584, \sii6716, 6731 emission-line data cube, where they stand out clearly against both the dominant \hii region population and the widespread diffuse emission.  The intrinsic emission is 
then extracted accounting for the highly inhomogeneous emission-line ``background'' throughout the field of view. 
Collisional to recombination line ratios like \sii/\Ha, \nii/\Ha, and \oi/\Ha tend to increase when the background emission is corrected for. We find that many (but not all) sources detected with the principal component analysis have properties compatible with supernova remnants (SNRs). Applying a combined \sii/\Ha and \nii/\Ha classification criterion leads to a list of 59 sources with SNR-like emission lines. Many of them exhibit conspicuous spectral signatures of SNRs around 7300 \AA, and a stacking analysis shows that these features are also present, except weaker, in other cases.  At nearly 30 Mpc, these are the most distant SNRs detected by optical means to date. We further report the serendipitous discovery of a luminous ($M_V \sim -12.5$), blue, and variable source, possibly associated with a supernova impostor.
\end{abstract}



\section{Introduction}
\label{sec:Intro}

NGC 4030 is an Sbc spiral with multiple arms \citep{Buta+15}, inclined by 47$^\circ$ \citep{Crowther13} at a distance of 29 Mpc\footnote{This value is a compromise between the redshift-based distance of $27 \pm 2$ Mpc 
\citep[considering a recession velocity of $1826 \pm 26$ km$\,$s$^{-1}$ relative to the CMB;][and adopting $H_0 = 67.8$ km$\,$s$^{-1}\,$Mpc$^{-1}$]{2004AJ....128...16K}, and the $29.9 \pm 0.2$ Mpc Tully-Fisher distance derived by  \citet{2013AJ....146...86T}.}. \citet{Ganda+06,Ganda+07} analysed the properties and kinematics of the gas and stellar population in the context of the SAURON survey, finding regular rotation in both gaseous and stellar components, and a prevalent young, metal-rich stellar population. Its central region contains a  nuclear star cluster and also an unresolved X-ray source \citep{Seth+08}. The galaxy was the host of SN 2007aa. As such it appears in several studies about the relation between SNe and their host galaxies \citep[e.g.][]{Smartt+09,Chornock+10,Anderson+12,Im+19}.

NGC 4030 is one of the star-forming galaxies targeted by the MUSE Atlas of Disks (MAD) survey \citep{Erroz-Ferrer+19,denBrok+20}, where its effective radius, stellar mass, star formation rate, and gas phase metallicity are estimated to be 31 arcsec (4.4 kpc), $10^{11.2} \mathrm{M_\odot}$, $11\, \mathrm{M_\odot\, yr^{-1}}$, and $12 +\log {\rm O/H} \sim 9.0$, respectively. More recently, these same data were part of the study by
\citet{Lopez-Coba+20} on outflows in MUSE galaxies, although no outflow was found for NGC 4030.

\begin{figure*}
  \includegraphics[width=2\columnwidth]{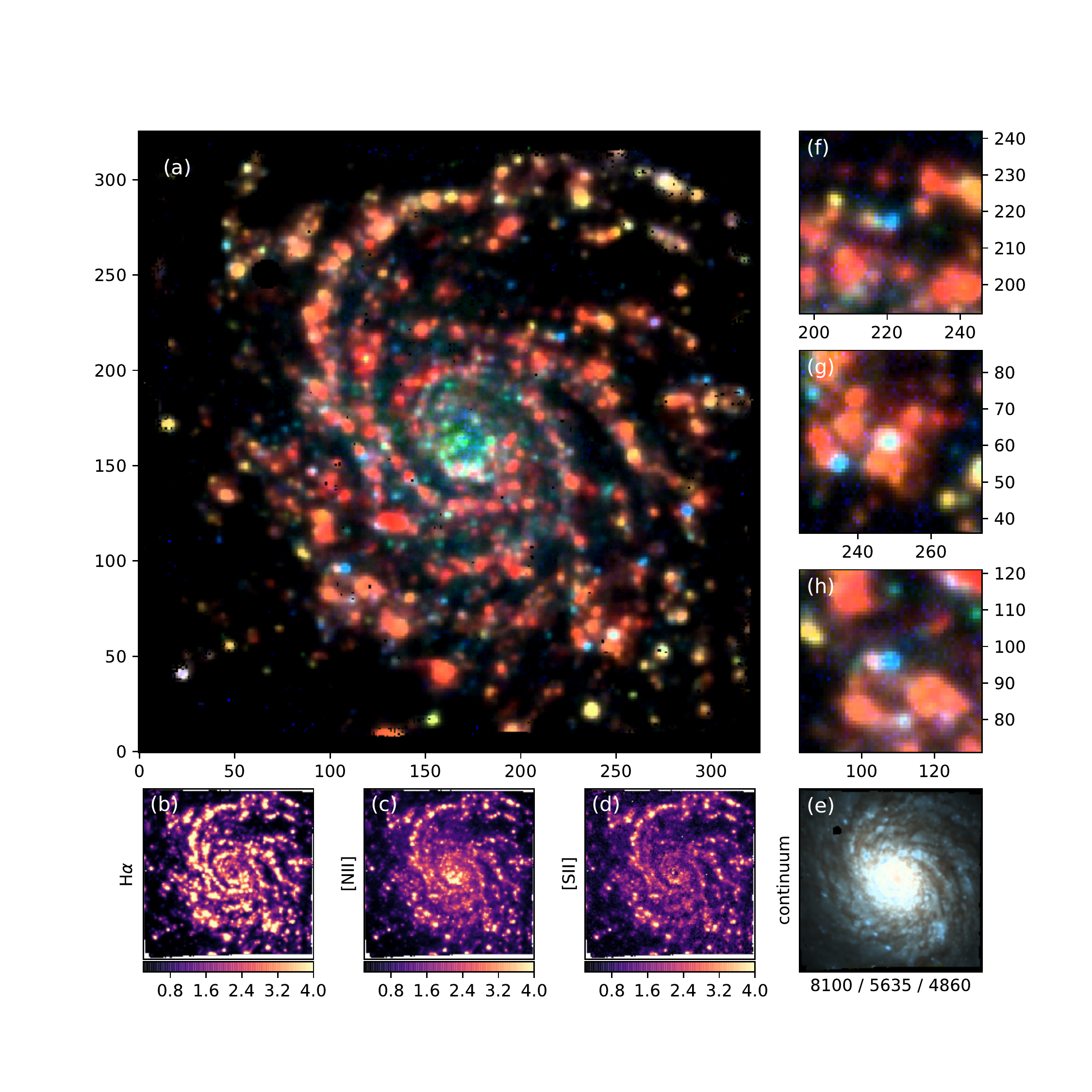}
  \caption{
{\em Bottom panels  (b, c, d)}: Maps of the \Ha, \nii 6584, and \sii 6716+6731 emission in NGC 4030. The intensity scale is given in units of the median spaxel luminosity (421,  164, and  94 $L_\odot$ for \Ha, \nii, and \sii, respectively).
{\em Panel (e)} shows an RGB composite based on the mean continuum flux densities at  $\lambda = 8100  \pm 50$  (red), $5635 \pm 45$ (green), and $4860 \pm 60$ \AA\ (blue, but excluding the $\pm 25$ \AA\ around \Hb). {\em Main panel:} RGB composite mixing \Ha (in red), \nii (green), and \sii + \oiii (blue) images as explained in the text. {\em Right-hand panels {(f, g, h)} :} Zoomed-in $50 \times 50$ spaxels cut-outs of panel (a) highlighting some of the blue/green compact sources we seek to identify.
}
\label{fig:RGBlines}
\end{figure*}

In this paper, we explore these same MUSE data, but with a very different focus. Inspection of emission-line maps reveals a population of compact sources whose pattern of line emission differs from that of  the  plethora of \hii regions throughout the disc of NGC 4030. The sources that caught our eye are seen in Fig.\ \ref{fig:RGBlines}, whose main panel shows a RGB composite where red codes for \Ha, green for the \nii line at 6584 \AA,  and blue for a combination of \sii6716, 6731 and \oiii 5007. Besides the impressive collection of \hii regions (in red), this image shows numerous blue/green point-like sources scattered all over the galaxy. 

What are these sources?  Are they planetary nebulae (PNe), like those found in MUSE data for NGC 628 by \citet{Kreckel+17}, supernova remnants (SNRs) like the ones recently found in NGC 6946 by \citet{Long+19}, or some other kind of source? Though the title of the paper spoils the suspense, let us pretend we do not know the answer and report the steps that lead to it, mimicking our own discovery process.

In order to identify the nature of these sources one first needs to devise ways of detecting them and quantifying their observational properties. These are the two central goals of this paper, which is organized as follows: Section \ref{sec:Data} describes the data and the processing steps. Section \ref{sec:EmLineAnalysis} presents a spaxel-based emission-line analysis using both conventional tools and principal component analysis (PCA). Section \ref{sec:SrcDetectionAndExtraction} describes how we detect and extract the observational properties of the sources. It also contains the first direct evidence that at least some of them are SNRs. This is further explored in Section \ref{sec:Results},  where we tailor the search for SNRs by means of emission-line-based criteria. Section \ref{sec:Discussion} discussed what impact these results might have on studies of star-forming regions in galaxies. Our main findings are summarized in Section \ref{sec:Conclusions}.

\section{Data and processing steps}
\label{sec:Data}

NGC 4030 was observed with the Multi Unit Spectroscopic Explorer (MUSE) at ESO's Very Large Telescope as part of the MAD survey \citep[PI: Carollo; see][]{Erroz-Ferrer+19}. Two 1800 s observations taken in February and March of 2016 were combined into a single $326 \times 326$ datacube with a spatial sampling of 0.2 arcsec covering a field of view (FoV) of  $\sim 1$ arcmin$^2$, and a spectral sampling of 1.25 \AA\ from 4750 to 9350 \AA.  The reduction procedures are those from the MUSE pipeline \citep{Weilbacher+12}. The seeing full width at half-maximum (FWHM) was 0.6 arcsec. At $d = 29$ Mpc, each spaxel covers 28 pc, and the seeing disc spans 84 pc.

The data cube went through our full analysis ``pipeline'', comprising basic pre-processing steps, fitting of the stellar continuum, and measurement of the emission lines. We refer the reader to \citet{ValeAsari2019}, where these same steps were applied to MaNGA galaxies, for details. Briefly, after masking foreground stars and low signal-to-noise spaxels, correcting for Galactic extinction\footnote{We used a Galactic $E(B-V) = 0.023$ \citep[from][]{Schlafly+2011} and a \citet{Cardelli+89} extinction law with $R_V = 3.1$.}, rest-framing and re-sampling the spectra, each spaxel had its spectrum fitted with the spectral synthesis code {\sc starlight} \citep{CF05} in order to map the stellar populations of the galaxy, and, more importantly for this work, to aid the measurement of nebular emission. 
The {\sc starlight} model is then subtracted from the observed spectrum, leaving a ``pure emission-line'' residual  spectrum $R_\lambda$ from which emission lines are measured by fitting Gaussians.

This pipeline produces a series of stellar population and nebular properties, but this study relies almost exclusively on the emission line fluxes. The main set of lines consists of \Hb, \oiii5007, \Ha, \nii6584, \sii6716,  and 6731, as these are measurable throughout most of the data cube. The median signal-to-noise ratio for these lines over the FoV are 26, 8, 115, 45, 15, and 11, respectively. Many other transitions are detected as well, as will be shown below.  Indeed, these fainter lines play an important role in this paper.


\section{Emission-line analysis}
\label{sec:EmLineAnalysis}

We start our analysis by examining the emission-line maps obtained from our pipeline (Section \ref{sec:EmLineMaps}), as this is how the sources that motivate this study were originally spotted. We then move to a less conventional space where individual lines are replaced by eigen line spectra (\ref{sec:PCAlines}), a change of variables which proved  useful in the context of this paper.

\subsection{Emission-line maps}
\label{sec:EmLineMaps}

The maps  at the bottom of Fig.\ \ref{fig:RGBlines} (panels b, c, and d) show the \Ha,  \nii and \sii  images obtained from our pipeline. The RGB composite (panel a) was built as follows: The red channel corresponds to the \Ha flux in units of its median value among all $93456$ usable spaxels. For the green one we multiply this same \Ha flux by $(\nii/\Ha) / 0.39$, where 0.39 corresponds to the median value of the \nii/\Ha ratio in the FoV. The blue channel is built analogously, but using the average of  $(\sii/\Ha) / 0.22$ and $(\oiii/\Hb) / 0.26$, where \sii denotes the sum of the 6716 and 6731 lines. 
These median line ratios lie in between those typical of metal rich \hii regions and those found in 
low-ionization nuclear emission-line regions (LINERs) and in the diffuse ionized gas \citep[see, for instance,][]{Lacerda+2018}. \hii regions thus appear in red in our colour scheme, while spaxels with larger collisional to recombination line flux ratios are painted in shades of green and blue -- \citet{Sanchez+20} and \citet{Lopez-Coba+20} employ similar color schemes.

\hii regions dominate the line emission everywhere in NGC 4030 except for the inner regions, where, as is common in bulges, the pattern of line emission is typical of LINERs. Its median $\nii/\Ha$ of 0.74 (nearly twice the value for the whole FoV) within 2 arcsec of the nucleus is what gives its green look in Fig.\ \ref{fig:RGBlines}. We note in passing that  there is no evidence of nuclear activity in this galaxy \citep[but see][]{Seth+08}. In fact, we find that its central regions are well within the ``retired galaxy'' regime  \citep{Stasinska+08,CF11}, with an \Ha equivalent width of just 1.4 \AA.  In any case, to avoid confusion we exclude the inner 4 arcsec in diameter from the analysis that follows.

The sources we are interested in are the blue/green dots scattered throughout the galaxy. 
The zooms in panels f, g, and h of Fig.\ \ref{fig:RGBlines} show a few examples. 
Close inspection of these images corroborates the visual impression that these are indeed unresolved, with sizes compatible with the ${\rm FWHM} = 0.6$ arcsec seeing. From here on we will refer to these sources as UFLOs, for unidentified forbidden line objects. The central goal of this paper is to remove the ``U'' from this acronym for as many sources as possible. From the very title of this paper one already knows that we claim that SNRs are amongst these UFLOs. We nonetheless keep this nomenclature until this is properly demonstrated, and also because this identification may not apply to all cases.

As also seen in these images, the same pattern of line emission also appears in diffuse form, both in the bulge and inter-arm regions, whose colours are similar to those of the unresolved UFLOs we aim to study. Whether this indicates that at least part of this diffuse emission comes not from genuinely diffuse ionized gas (DIG), but from a sparse collection of  sources similar to the ones we see in Fig.\ \ref{fig:RGBlines}, except weaker, is an interesting question, but one that will not be addressed here. As stated before, we seek to properly isolate and inspect the properties of the UFLOs in NGC 4030.

Fig.\ \ref{fig:RGBlines} encapsulates the rationale behind classical diagnostic diagrams involving collisional to recombination line flux ratios \citep[e.g.][hereinafter BPT]{BPT81}. 
We refer the reader to \citet{Lopez-Coba+20}, where several such diagrams are presented for NGC 4030 and many other galaxies. Red spaxels in Fig.\ \ref{fig:RGBlines} have line ratios typical of metal rich \hii regions, while green/blue spaxels occupy loci in diagnostic diagrams which overlap with those of SNRs, PNe, active galactic nuclei, and DIG-like emission.

The latter two of these  possibilities can be discarded right away, since (1) UFLOs appear all through the disc of NGC 4030, and (2) they are compact. PNe can also be ruled out on the grounds that, besides not being  strong \oiii emitters, our UFLOs are far too luminous to be associated with PNe, whose \Ha luminosities range from $\sim 1$ to  100 $L_\odot$ \citep{Stasinska+98}.
This is much fainter than even the spaxel luminosities of our UFLOs, which (due to atmospheric seeing) account for just part of the total output of any unresolved source.  PNe are just too faint to be detected in these data.
SNRs, on the other hand, remain a viable possibility. With expected diameters of the order of 40 pc \citep{Roth+2018}, SNRs would appear unresolved at the 84 pc resolution of the MUSE data.

\subsection{PCA tomography}
\label{sec:PCAlines}

Let us now explore a less canonical approach, where the choice of line properties and diagrams to use is guided not by pre-conceived ideas on how to best combine emission-line data to gain physical insight, but by algebra.

\begin{figure}
  \includegraphics[width=\columnwidth]{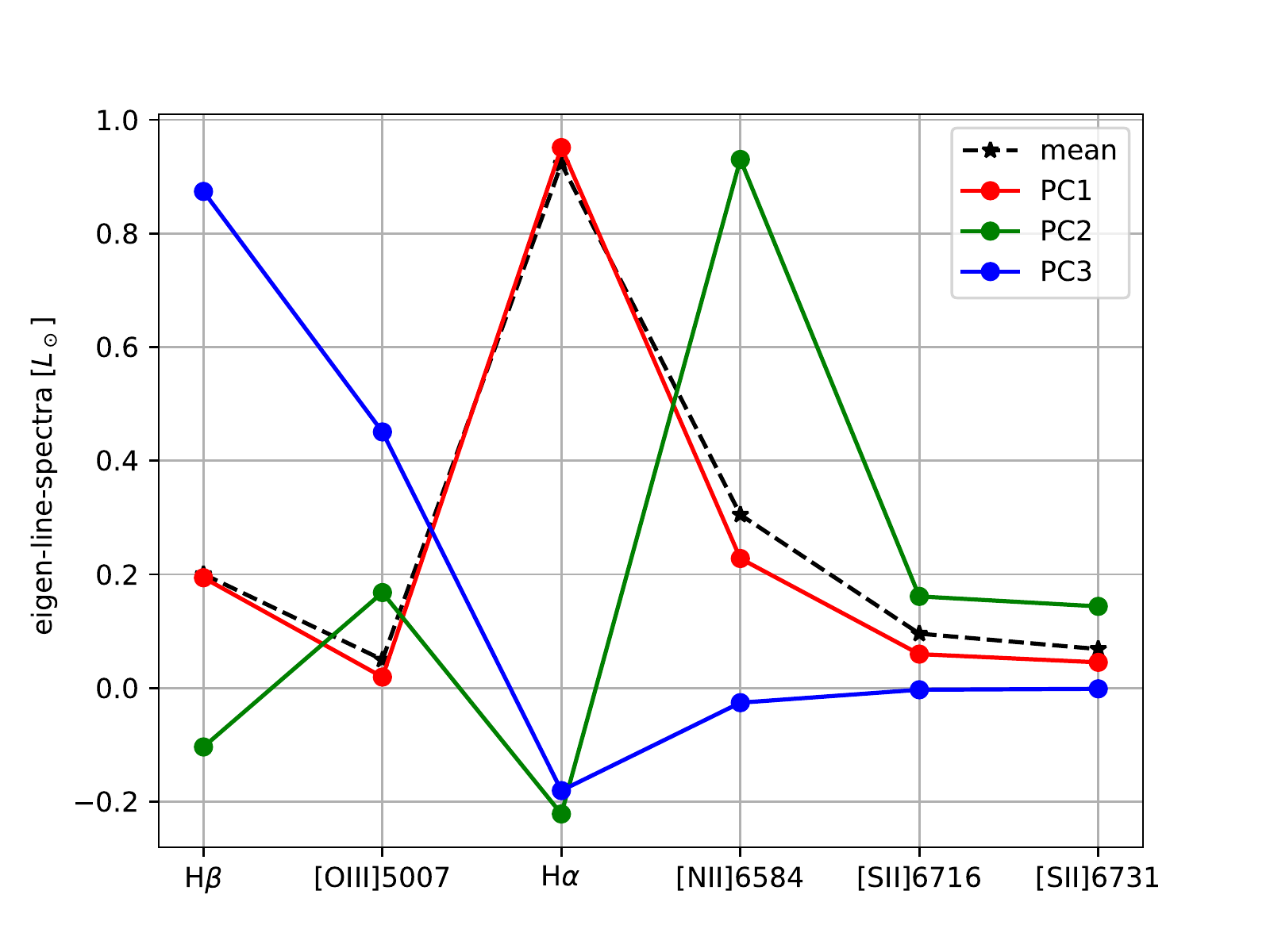}
  \caption{
The first three eigen line spectra obtained from the PCA. The dashed line shows the rescaled mean line-spectrum.
}
\label{fig:eigSpec}
\end{figure}

\begin{figure*}
  \includegraphics[width=2.12\columnwidth, trim={0 1cm 0 0}]{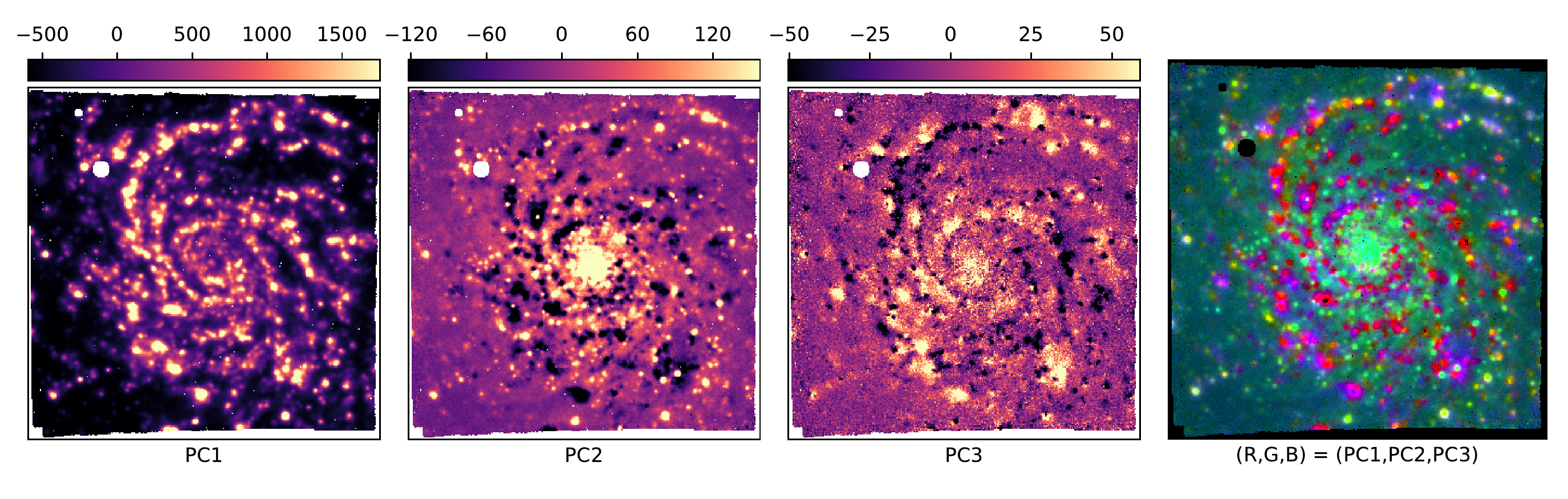}
  \caption{
Tomograms of the first three principal components. The intensity scales (all in $L_\odot$) were set to span the range from the 2.5 to the 97.5 percentile of the PC values in each image.
Missing points (in white) mark the masked spaxels (bad data or foreground stars).
The image on the right shows a composite mapping PC1, PC2, and PC3 (each rescaled to its 2.5--97.5 percentile range) to red, green, and blue, respectively.
}
\label{fig:Tomograms}
\end{figure*}

Inspired by the pioneering work of \citet{Steiner+09}, we performed a series of PCA tomography experiments with the MUSE data on NGC 4030. Here we focus on the results obtained for a $ 326 \times 326 \times 6$ cube where the six layers correspond to the luminosities of \Hb, \oiii5007, \Ha, \nii6584, \sii6716, 6731. The analysis is carried over the spaxels where all six lines were detected by our pipeline, regardless of signal-to-noise. The terminology ``line-spectrum'' will be used to denote a set of these six emission-line luminosities.\footnote{PCA tomography was also applied to the full data cube, but in this paper we focus on the $326 \times 326 \times 6$ emission-line data cube only. Incidentally, a faint foreground star at pixel $(x,y) = (46, 302)$ was first identified by us in a full spectrum PCA, and masked from the analysis thereafter.}

We obtain that the first, second and third PCs account for 98.5, 1.2, and 0.2 per cent of the total variance, respectively. Fig.\ \ref{fig:eigSpec} shows the first three eigen line spectra derived from the data (in red, green and blue lines, respectively), along with the mean line spectrum (shown as a black dashed line, scaled to unit norm). Unsurprisingly, this mean line-spectrum is typical of the \hii regions in NGC 4030. Moreover, the first eigen line spectrum is very similar to the mean line-spectrum, which is just what one expects since the spaxel line-spectra were not rescaled prior to the PCA. The main role of PC1 is thus to set the overall flux scale. 

The second principal component has an eigen line spectrum that contrasts forbidden (\oiii, \nii, and \sii)  with recombination (\Ha, \Hb) lines. Spaxels with positive values of PC2  map into those with enhanced forbidden line emission, like the blue/green spaxels of Fig.\ \ref{fig:RGBlines}a. Finally, the third eigen line spectrum essentially boosts \Hb (and \oiii) with respect to \Ha. We attribute this behaviour to the effects of dust. Spaxels with PC3 $> 0$ have larger \Hb/\Ha and are thus presumably less attenuated by dust, and vice versa.

Attributing meaning to PCs is often a dangerous exercise. In the case of  data cubes, however, the spatial organization of the PCs gives valuable insight on the astrophysical meaning of this otherwise purely mathematical construct. This is the beauty of PCA tomography.

Fig.\ \ref{fig:Tomograms} shows tomograms of the first three PCs. Tomogram 1 evidently traces the \hii regions in the galaxy. In fact, this image of PC1 is very similar to the \Ha map in panel b of Fig.\ \ref{fig:RGBlines}. Tomogram 2, on the other hand, highlights regions of enhanced forbidden line emission. \hii regions appear as dark zones, while UFLOs stand out, along with faint diffuse emission both in the central and inter-arm regions. 
In tomogram 3 the spiral arms alternate from positive to negative values of PC3 as one  crosses them in azimuth. It is also visible that $<0$ regions are more concentrated that $>0$ ones. 
Inspecting the image one further finds several dark (PC3 $< 0$) spots surrounded by bright ones (PC3 $> 0$), but not the other way around. Recalling that negative PC3 maps to an increase in \Ha/\Hb, both these facts are consistent with our association of PC3 with dust. 

The rightmost panel of Fig.\ \ref{fig:Tomograms} combines tomograms 1, 2, and 3 into a RGB composite. 
\hii regions appear as mixtures of red (PC1) and blue (PC3). UFLOs stand out as compact green/yellow sources. These same colours also appear in a diffuse component all over the FoV, brighter in the central regions.

Given its association with UFLOs, PC2 is the most relevant component in the context of this paper. Accordingly, in the next section we make use of tomogram 2 to select sources to study more closely.


\section{Detecting and extracting the sources}
\label{sec:SrcDetectionAndExtraction}

This section presents our efforts to produce a list of candidate UFLOs in NGC 4030, and examine their properties. Section \ref{sec:Detection} describes the source detection method, while in Section \ref{sec:Background}  we deal with the difficult problem of how to isolate them from the intense and inhomogeneous background they are immersed in. A diagnostic diagram analysis is also presented, which suggests a link between UFLOs and SNRs. Spectral extraction is dealt with in Section \ref{sec:SpectralExtraction}, which also presents direct evidence that many of our UFLOs are in fact SNRs. Section \ref{sec:TheBeast} presents our star case of this association.

\subsection{Detecting UFLOs}
\label{sec:Detection}

\begin{figure}
  \includegraphics[width=\columnwidth]{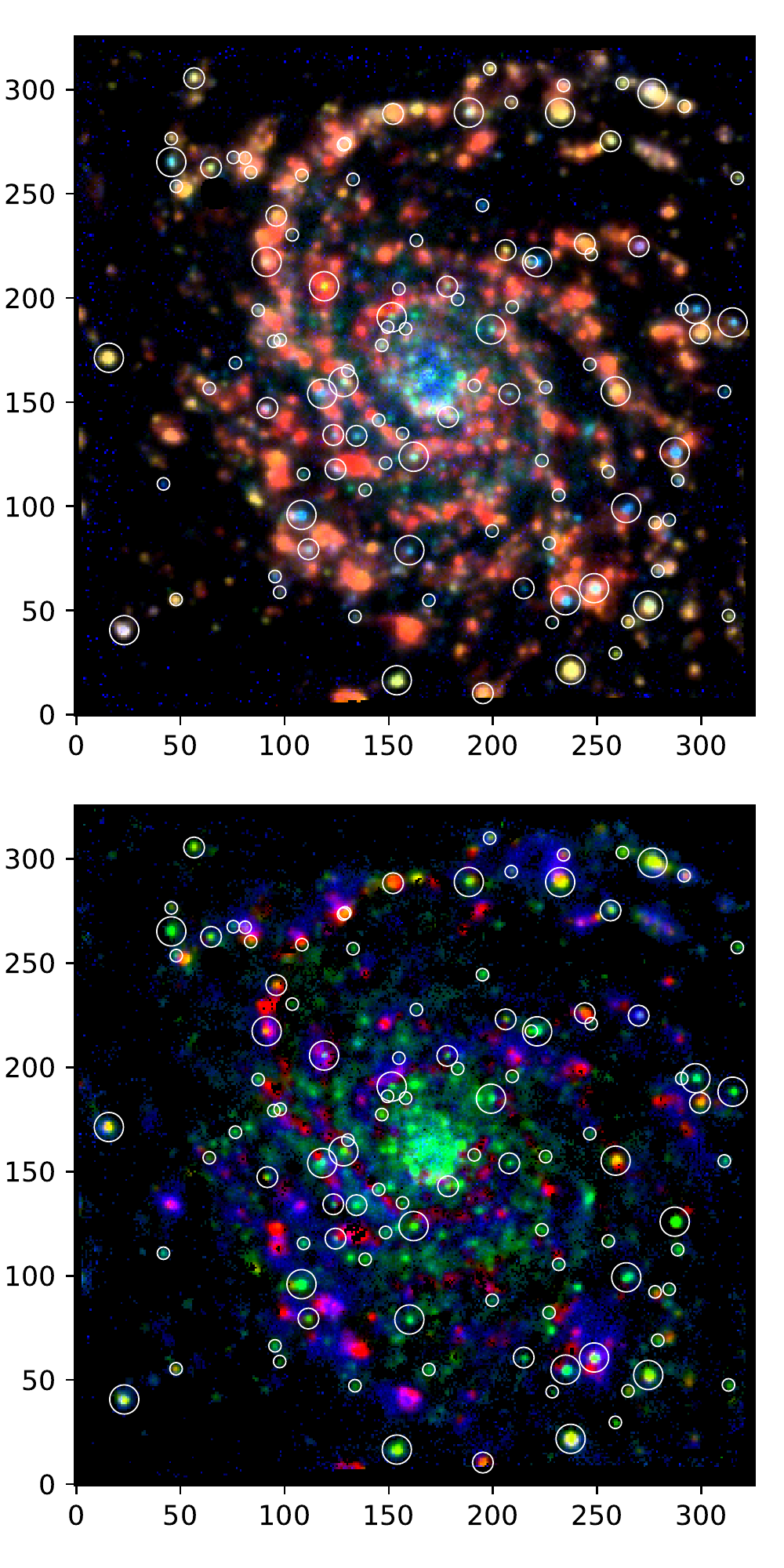}
  \caption{RGB composites like those in Figs.\ \ref{fig:RGBlines} (top) and  \ref{fig:Tomograms} (bottom),
but re-processed to darken the diffuse emission, marking the loci of candidate UFLOs. Large circles mark the 26 sources in the primary sample. Small and intermediate size circles are sources also detected for threshold values of 
 $T = 100$ and $200 L_\odot$, respectively.
}
\label{fig:SrcDetector}
\end{figure}

Our strategy to identify UFLOs is to search for unresolved sources in the PC2 image. We use the astropy implementation of DAOStarFinder on tomogram 2 after masking spaxels with PC2 $< 0$  and excluding regions within 2 arcsec from the nucleus. The expected FWHM of the sources is fixed at three spaxels (the seeing). The DAOStarFinder threshold parameter ($T$) is first set to $T = 300 L_\odot$. This configuration produces a list of 27  candidate UFLOs. Naturally, lower threshold levels lead to more detections: 53 sources for $T = 200 L_\odot$  and 147 for $T = 100 L_\odot$. 

Fig.\ \ref{fig:SrcDetector} marks the detected sources on the RGB composites of Figs.\ \ref{fig:RGBlines} (top) and  \ref{fig:Tomograms} (bottom), both of which (especially the bottom one) were modified to darken the diffuse emission in order to facilitate the visualization of the sources. 
Candidates detected with ${\rm PC2} > T = 100$, 200, and $300 L_\odot$ are marked with circles of radius  3, 5 and 7 pixels, respectively, but only sources that satisfy the criteria outlined in the next section are actually marked. In order to focus first on the clearest detections, we shall treat the sources obtained with $T = 300 L_\odot$ as our primary UFLO sample.

\begin{figure}
  \includegraphics[width=\columnwidth]{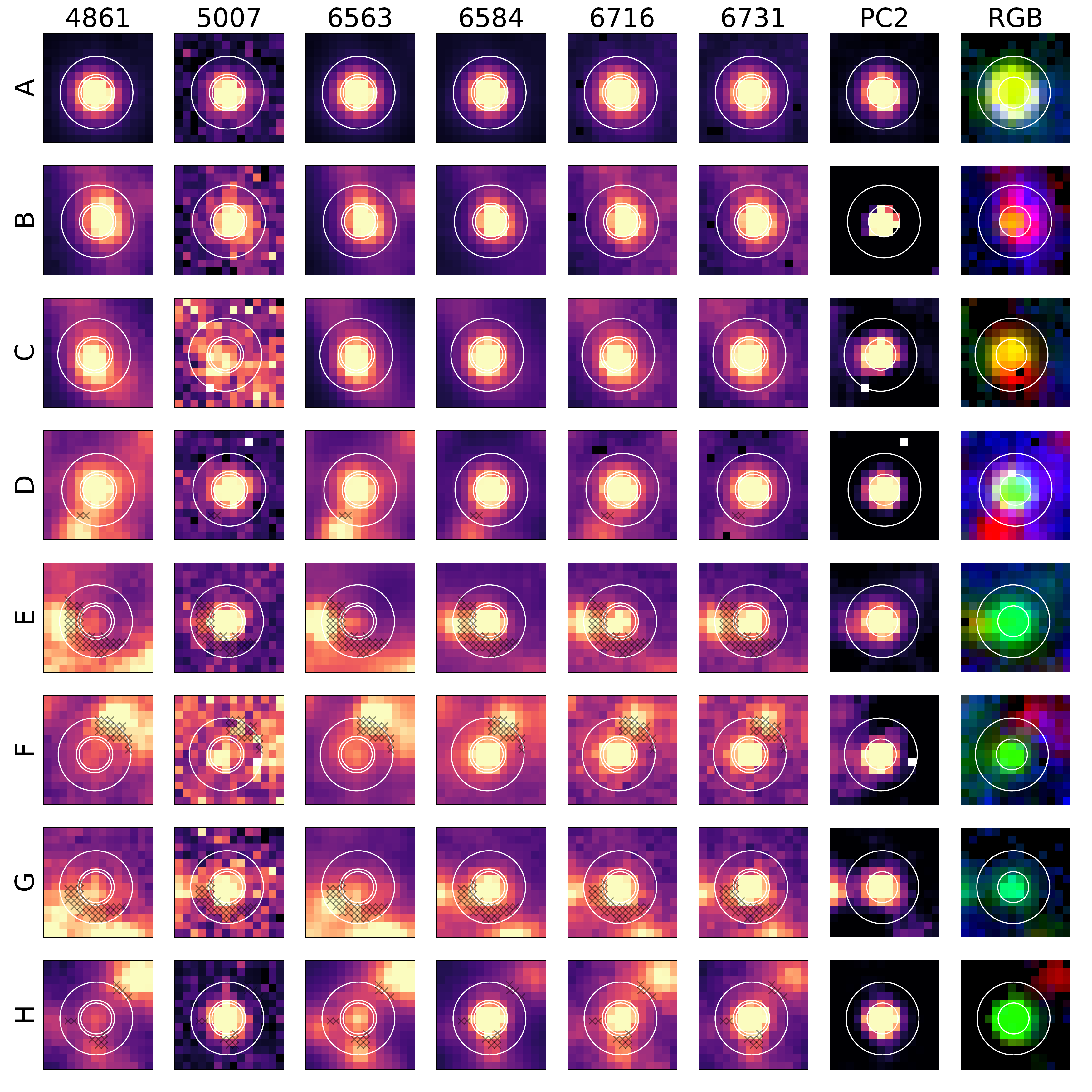}
  \caption{
 $18 \times 18$ spaxels cut-outs centred on a few example candidate UFLOs.
Columns show the images in (from left to right) \Hb, \oiii, \Ha, \nii, \sii 6716, \sii 6731, and tomogram 2. The intensity scale in each map goes from 0 to its 95th percentile level. The rightmost panel shows the RGB composite of tomograms 1, 2, and 3. Circles with radii of $r_{\rm src} = 2.55$, $r_{\rm in} = 3$, and $r_{\rm out} = 6$ spaxels are drawn. Crosses mark spaxels between $r = 3$ and 6 that are brighter in \Ha than the mean \Ha surface brightness within $r < r_{\rm src}$ source aperture.
}
\label{fig:BckExamples}
\end{figure}

With the detection method established, we now need to have a closer look at each source and verify whether its properties can be reliably extracted. This bring us to the background problem.

It is critical to realize that Fig.\ \ref{fig:SrcDetector} gives the false impression that UFLOs are easily separable from the surrounding emission. This impression comes about because of the image processing done to suppress diffuse emission and especially \hii regions. Many of our candidate sources are much harder to spot in the original images, particularly those in \Ha and \Hb. To illustrate this, Fig.\ \ref{fig:BckExamples} shows cut-outs around some of the detected sources for our six main set emission lines, as well as for tomogram 2 and the PC RGB composite. The examples shown span cases where the source is clearly detected in all lines to others where it is only visible in forbidden lines (sometimes barely so in \oiii). The environment of the sources also varies, with some sitting in relative isolation but most having intense line emission within a few spaxels of the source. The structure in this  ``background'' is more clearly seen in the \Ha and \Hb panels, where, unlike for the forbidden lines, the brightest spaxels within the cut-out area are often not on the source itself. These issues become more severe for samples defined with lower $T$ values, as exemplified by cases G and H in Fig.\ \ref{fig:BckExamples}, which (unlike the others in this plot) are not part of our primary sample.

None of this invalidates our detection method. In fact the tomogram 2 cut-outs in Fig.\ \ref{fig:BckExamples} always show a cleanly isolated source, which is what makes it an excellent detection image. Yet, these examples show that extracting the source properties from such an inhomogeneous and bright background is a complex task. To produce a workable sample of UFLOs we must therefore first implement a method to extract their properties.

\subsection{Source extraction}
\label{sec:Background}

Our first attempt to extract the line luminosities of the sources was to perform a simple aperture photometry, with the source contained within a circle of radius $r_{\rm src}$, and the background level defined over an external annulus between $r_{\rm in}$ and $r_{\rm out}$. For the source aperture we use $r_{\rm src} = 2.55$ spaxels (corresponding to $2 \sigma$ of a gaussian of FWHM $= 3$), while the background region is defined by $r_{\rm in} = 3$ and $r_{\rm out} = 6$. The astropy.photoutils package was used for this task.

As one could anticipate from Fig.\ \ref{fig:BckExamples}, this simple scheme often fails for the Balmer lines because the background level outshines the source itself. In some cases this also happens for forbidden lines (particularly \oiii, which is generally weak in NGC 4030). Alternative methods involving the fitting of complex 2D surfaces were tried (e.g., a gaussian point source plus a 2D Legendre polynomial to represent the background), with limited success. 

After experiments, we settled for a recipe that removes from the background all spaxels that are brighter in \Ha than the mean \Ha\ brightness within the source aperture. Examples of such masked spaxels are marked with a cross in the cut-outs in Fig.\ \ref{fig:BckExamples}. Cases E, F and H, for instance, have a substantial fraction of the spaxels between $r_{\rm in}$ and $r_{\rm out}$ masked, while the others have none or just a few.  A second (and less important) adjustment in the method is that in the computation of the background level we now give less weight to spaxels that are farther away from the source. We stress that this ad hoc extraction method is not meant to be optimal in any sense, but just a simple workaround to deal with the complex background landscape in NGC 4030. While it is likely that more sophisticated tools developed to deal with crowded field spectroscopy \citep[e.g.,][]{Fabrika+2005,Lehmann+2005,Kamann+2016,Roth+2018} can be used or adapted to this case, the method outlined above suits our needs for the current paper.

After applying this revised extraction method to our candidate sources, we remove from the list those where more than half of the background spaxels had to be masked for being too bright in \Ha. We further reject sources where any of the six emission lines in our main set comes out with negative flux after background subtraction. Of the 27 sources in our primary sample, only one is rejected by these cuts.
The survival fractions are smaller for samples culled with lower $T$ values. For $T = 200$ (100) $L_\odot$ the number of sources goes from 53 (147) to 45 (106).

\begin{figure}
  \includegraphics[width=\columnwidth]{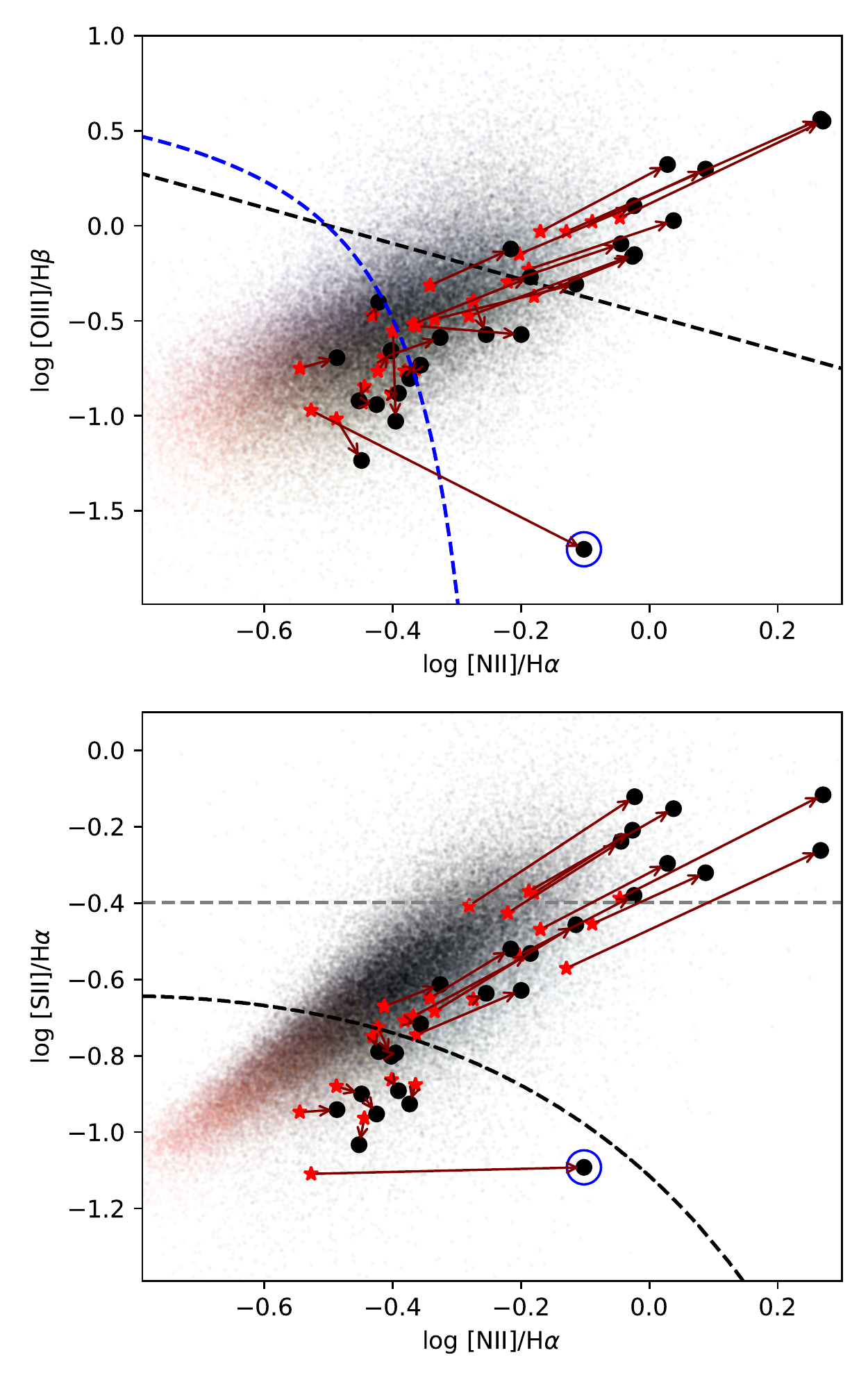}
  \caption{Diagnostic diagrams before (red stars) and after (black circles) background subtraction for the 26 sources in our primary UFLO sample (see text). The point marked with a blue circle is the peculiar source discussed in Section \ref{sec:TheWeirdo}. Dots correspond to individual spaxels in the whole FoV, colour-coded according to  the \Ha, \nii, \sii $+$ \oiii-based RGB composite in Fig.\ \ref{fig:RGBlines}. 
The BPT diagram (top panel) includes the \citet{Stasinska+06} demarcation line (in blue), while in the bottom panel the $\sii/\Ha = 0.4$ line (in gray) marks the conventional borderline to separate SNRs from \hii regions. Black dashed lines in both diagrams show the SNR/\hii region separation lines proposed by \citet{Kopsacheili+20}.
}
\label{fig:BPT4UFLOs}
\end{figure}

\subsubsection{Diagnostic diagrams}

Fig.\ \ref{fig:BPT4UFLOs} shows the \oiii/\Hb and \sii/\Ha vs.\ \nii/\Ha diagrams for the 26 sources in the primary sample. Red stars show the results obtained integrating all the line emission within the source aperture (i.e., without
background correction). Each star is connected to a black circle located at the ratios obtained after background subtraction. Ignoring sources that do not move much and the obvious outlier (marked with a blue open circle; 
see Section \ref{sec:TheWeirdo}), the tendency is for all line ratios to increase. This occurs because in most cases the ``background'' line emission is associated with \hii regions. Subtracting \hii region-like lines from sources that are defined (through PC2) in terms of excess forbidden line emission naturally leads to even larger collisional to recombination line ratios. Some of our UFLOs are thus even more different from \hii regions than initially thought.

Dashed lines in Fig.\ \ref{fig:BPT4UFLOs} mark borderlines for different kinds of sources in these diagrams. In  the BPT diagram (top panel) the blue line outlines the \citet{Stasinska+06} limit of pure star-forming galaxies, while the black line comes from the recent work by \citet{Kopsacheili+20}, who trace optimal demarcation lines on the basis of models for \hii regions and SNRs.  The bottom diagram includes the $\sii/\Ha = 0.4$ boundary commonly used to identify SNRs candidates \citep[e.g.][]{Blair&Long97,Blair&Long04,Long+19}. Some variation on this value exists in the literature; for instance, \citet{Dopita+10} and \citet{Leonidaki+13} use a less stringent $\sii/\Ha = 0.3$ criterion, while \citet{Matonick&Fesen97} use 0.45 instead, and \citet{Long+18} further note that the separation between \hii and SNRs becomes increasingly blurred at low surface brightness. \citet{Kopsacheili+20}  show that a simple $\sii/\Ha \geq 0.4$ demarcation line may induce biases against slow shocks.
Their optimal separator in the \sii/\Ha-\nii/\Ha space is shown as a black  dashed line, which runs well below the $\sii/\Ha = 0.4$ boundary.

Regardless of which demarcation line one choses to use, Fig.\ \ref{fig:BPT4UFLOs} shows that the background corrections make  some UFLOs line ratios more SNR-like. Focusing on the bottom plot, one sees that 9 UFLOs cross the $\sii/\Ha = 0.4$ boundary, while 16 fall into the SNR region as defined by \citet{Kopsacheili+20}.

\subsubsection{[O{\,\sc i}]$\lambda6300$}
\label{sec:OI6300}

\begin{figure}
  \includegraphics[width=1.11\columnwidth]{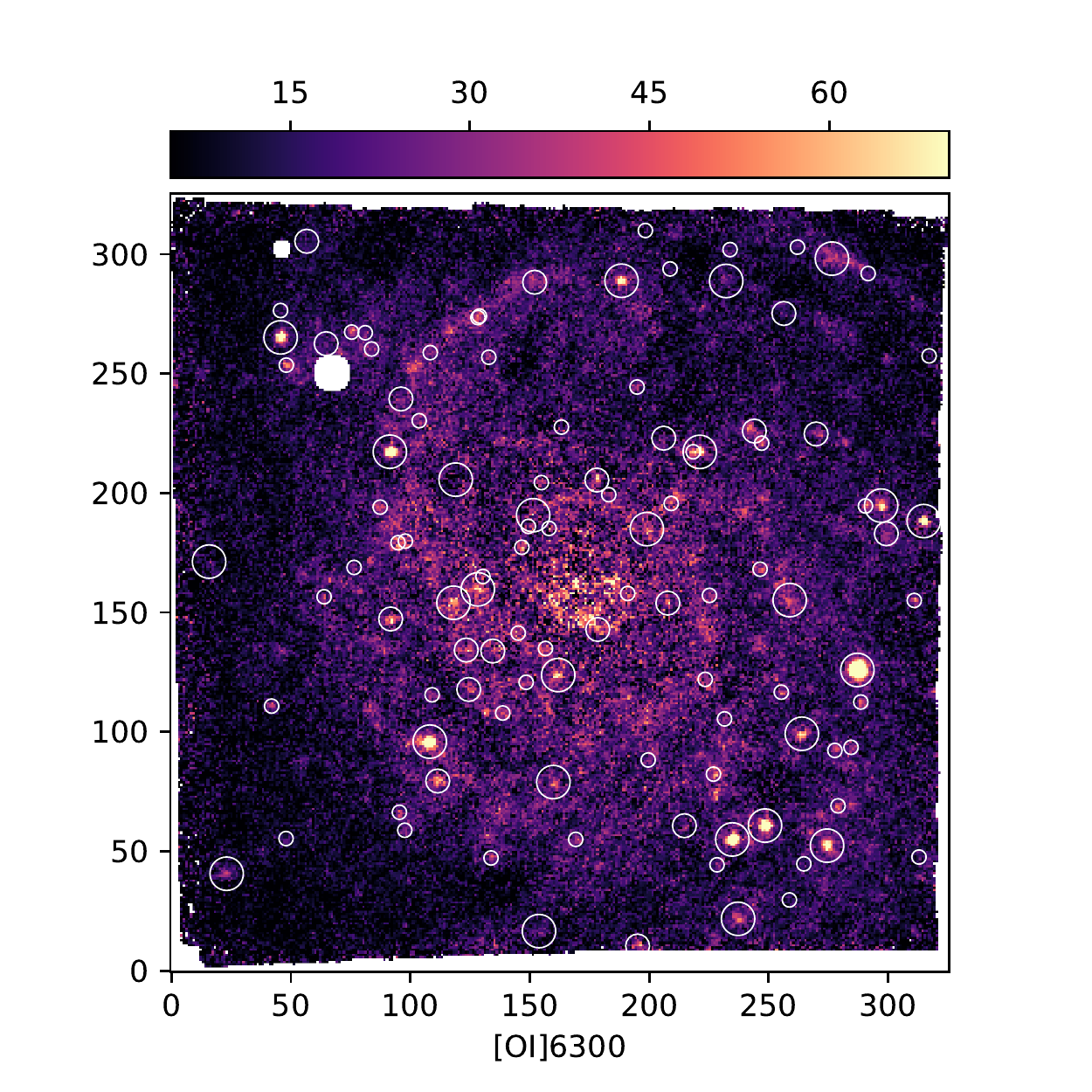}
  \includegraphics[width=\columnwidth, trim={0 1.5cm 0 0}]{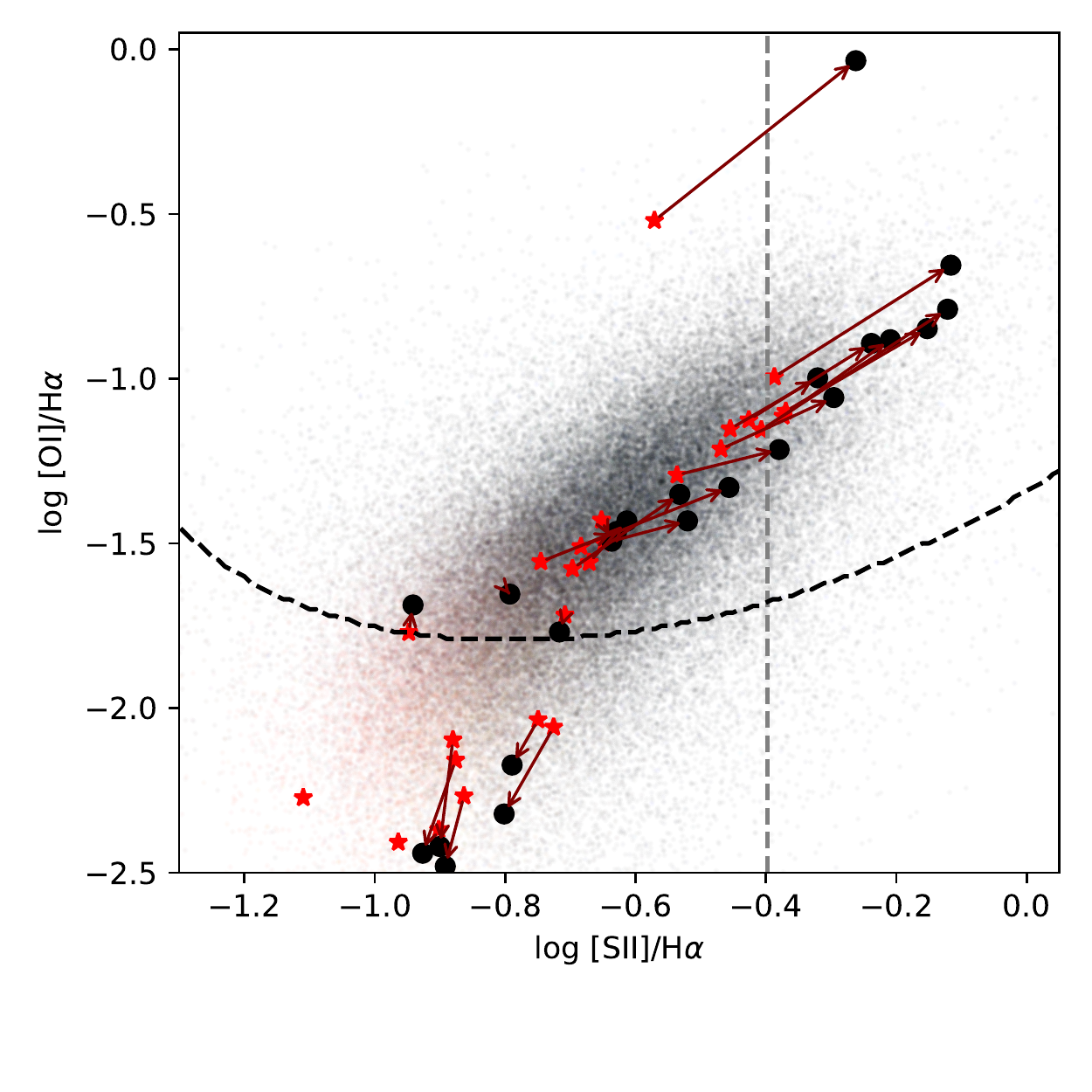}
  \caption{{\em Top:} Map of the [OI]6300 line luminosity (in $L_\odot$).  Circles mark UFLO detections as in Fig.\ \ref{fig:SrcDetector}. {\em  Bottom:} As Fig.\ \ref{fig:BPT4UFLOs}, but for the \sii/\Ha vs.\ \oi/\Ha diagram.
The conventional $\sii/\Ha = 0.4$  and the \citet{Kopsacheili+20} SNR/\hii region separation lines are shown in dashed gray and black, respectively. 
}
\label{fig:[OI]}
\end{figure}

The analysis presented so far has overlooked the \oi line at 6300 \AA, a well-known tracer of partially ionized zones, and thus a good indicator of the presence of shocks. This transition is usually very weak and hence prone to large measurement uncertainties, particularly in \hii regions, the dominant source of emission lines in NGC 4030. This is reflected on its median signal-to-noise ratio of just 3 over the FoV, compared to 8 for \oiii, the weakest of the six lines in our main set. Now that we have identified a population of sources likely associated with SNRs, let us examine the \oi data to verify whether they are indeed strong \oi emitters.

Fig.\ \ref{fig:[OI]} (top) shows the \oi map for NGC 4030. Primary sample UFLOs are marked with large circles. Intermediate and small circles are used for UFLOs in the  $T = 200$ and $100 L_\odot$ samples (as in Fig.\ \ref{fig:SrcDetector}). The source at pixel $(x,y) = (287, 126)$ stands out as the brightest \oi emitter in the galaxy -- this extreme case is further discussed in Section \ref{sec:TheBeast} (see also Fig.\ \ref{fig:TheBeast}). Several of the other primary sample UFLOs and many of those in the other samples are also coincident with peaks in the \oi image.

The bottom panel in Fig.\ \ref{fig:[OI]} shows the \oi/\Ha vs.\ \sii/\Ha  diagram. As in Fig.\ \ref{fig:BPT4UFLOs},
red stars and black circles mark the line ratios obtained prior to and after background subtraction, respectively. Most UFLOs line up along the highly correlated cloud of spaxel values. As for other forbidden lines, \oi/\Ha tends to increase with the background subtraction, populating the upper right part of the diagram. The plot also shows the conventional SNR/\hii division at $\sii/\Ha = 0.4$ as well as the optimal model separator line according to \citet{Kopsacheili+20}. Sources that lie above the Kopsacheili et al. line in the \sii/\Ha vs.\ \nii/\Ha diagram also 
lie above the corresponding line in this plot.

\subsubsection{Many (but not all) UFLOs have SNR-like line ratios}

The title of this subsection summarizes the result of the diagnostic diagram analysis in Figs.\ \ref{fig:BPT4UFLOs} and \ref{fig:[OI]}. On the one hand, the diagnostic diagrams presented above support the interpretation that many of our UFLOs are associated with SNRs. On the other, however, these same diagrams show that several of our sources remain within or close to the \hii region areas even after background subtraction. 

For instance, of the 11 sources in the bottom plot of Fig.\ \ref{fig:BPT4UFLOs} initially located within the star-forming zone as defined by the \citet{Stasinska+06} criterion (blue dashed line in the BPT diagram), 8 remain there after background subtraction, and similarly for Fig.\  \ref{fig:[OI]}.
These non-SNR sources have line ratios slightly offset from the main body of \hii region-dominated spaxels (red dots) in these diagnostic diagrams, which explains why they are picked by PC2.

\subsection{Spectral extraction}
\label{sec:SpectralExtraction}

\begin{figure*}
  \includegraphics[width=2\columnwidth]{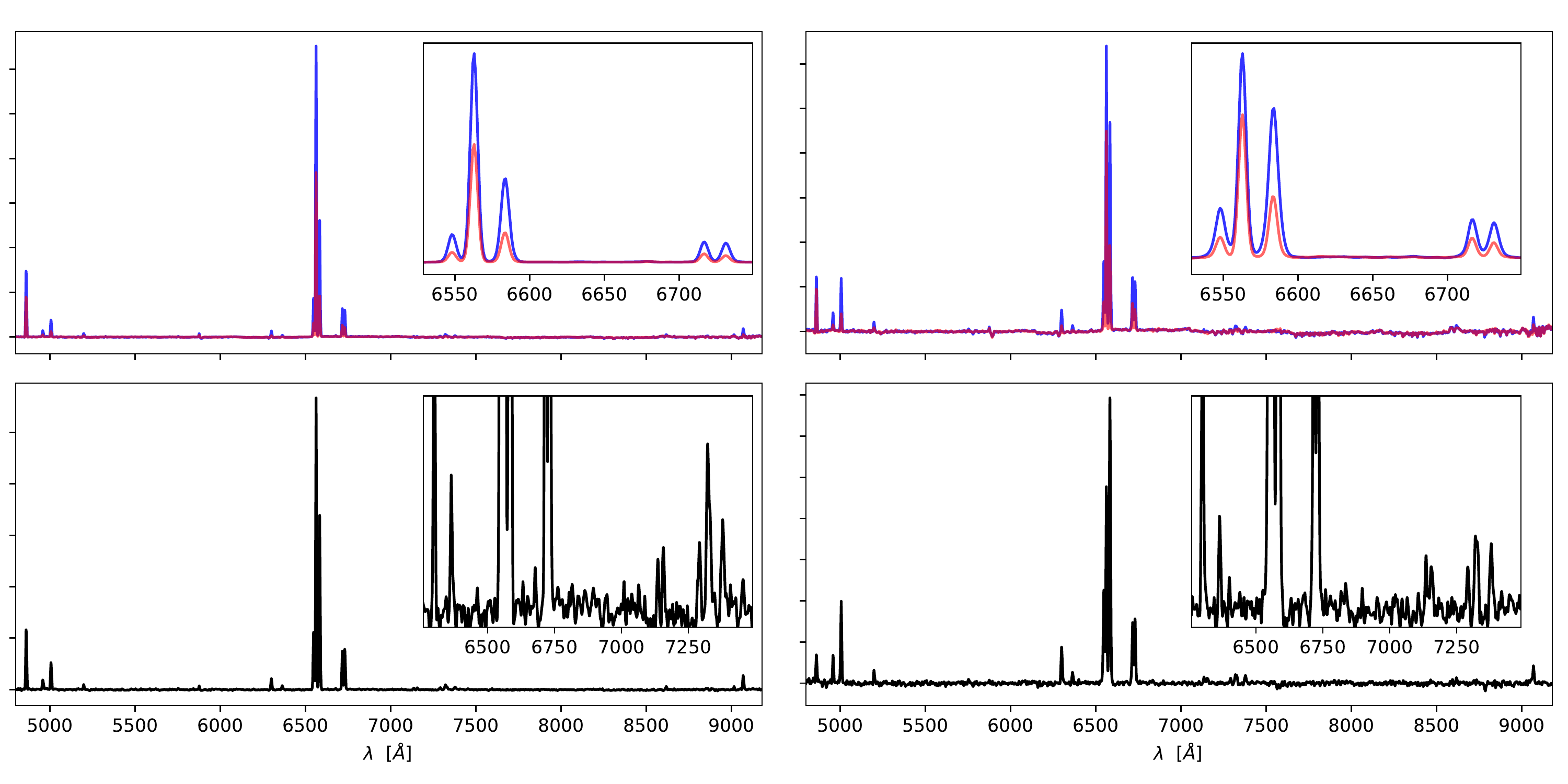}
  \caption{Examples of the spectral extraction for two sources. Top panels show the spectra in the source aperture (in blue) and its estimated background (red), with insets zooming on the \Ha, \nii, and \sii lines.
The source minus background difference spectra are shown in the bottom panels, with insets zooming in the 6250--7500 \AA\ window. All spectra were slightly smoothed (with a $\sigma = 2$ \AA\ gaussian) for visualization purposes.  
}
\label{fig:ExampleSpectralExtraction}
\end{figure*}

Up to this point we have worked exclusively the integrated emission-line measurements. Now that we have managed to produce a sensible list of UFLOs, let us examine their spectra. 

Just as with the line images, to extract a source's spectrum one needs to properly subtract its background. We do this in exactly the same way we did for the individual line images, 
except now operating on a $\lambda$ by $\lambda$ basis.
We perform the extraction on the residual $R_\lambda$ obtained after subtracting the {\sc \starlight} model fit from the observed spectrum, as this represents our best estimate of the pure emission-line spectrum. Prior to the spectral extraction the $R_\lambda$ for every individual spaxel is shifted to the rest frame using the \Ha-based velocities, thus eliminating possible effects of the galaxy's rotation. 

Fig.\ \ref{fig:ExampleSpectralExtraction} exemplifies the spectral extraction process for two UFLOs.
Blue and red lines in the top panels represent the spectra in the source ($r < r_{\rm src} = 2.55$) and background 
($3 < r < 6$)  apertures, respectively. The inset panels, which zoom into the \nii-\Ha-\sii window, show that (1)  the background contribution is significant in both cases, and that (2) the sources have visibly larger \nii/\Ha and \sii/\Ha ratios than their surroundings.

The intrinsic (i.e., background-subtracted) source spectrum is shown in the bottom panels. Weak lines such as \nOne 5199, \oi 6300, 6364, and \siii 9069 show up clearly even in these autoscale plots. The insets reveal several other features, the most striking of which are the lines around 7300 \AA. {\em These are telltale signs of SNRs.}

\subsection{The smoking gun}
\label{sec:TheBeast}

\begin{figure*}
  \includegraphics[width=2\columnwidth]{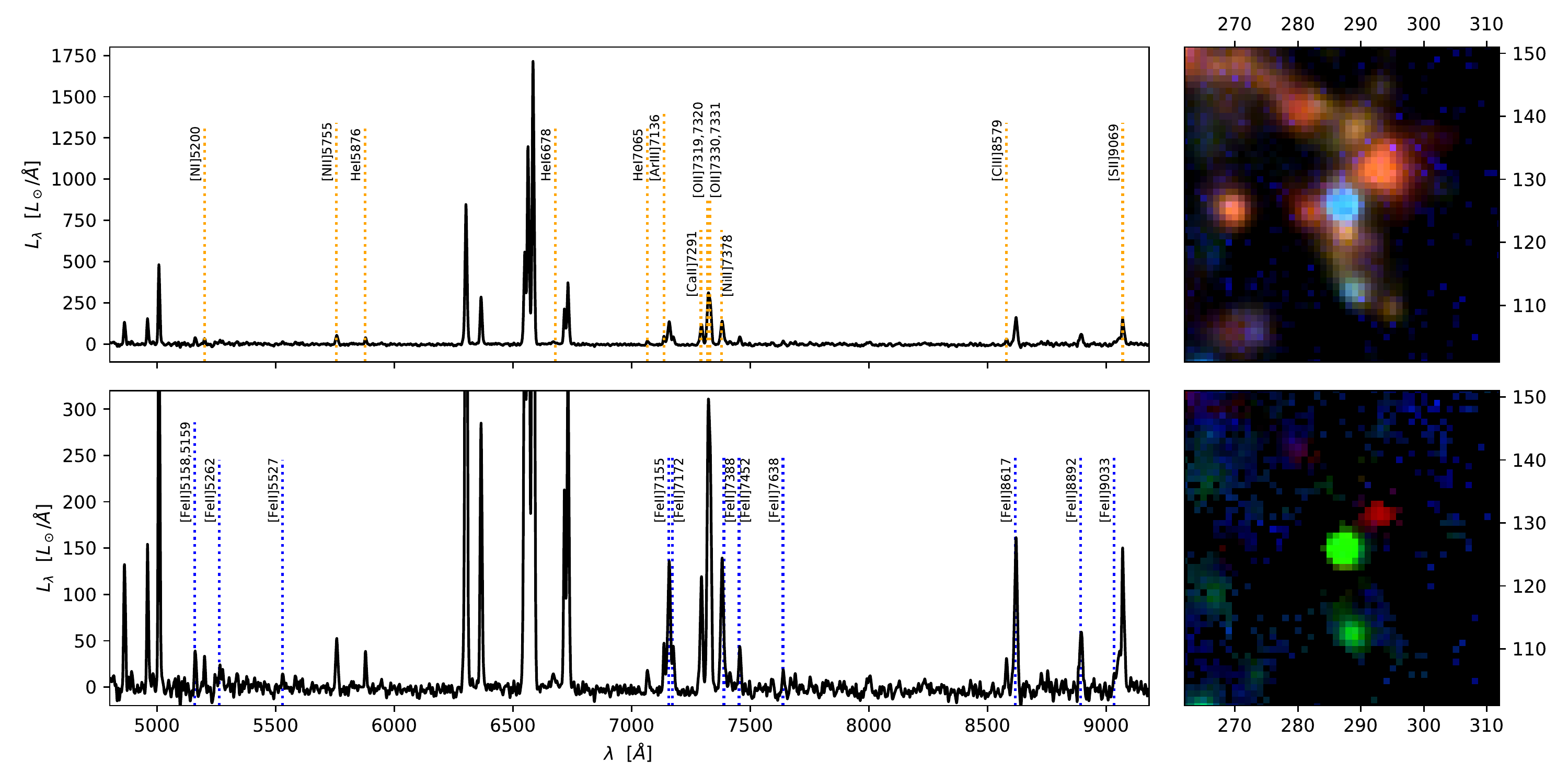}
  \caption{{\it Left:} Spectrum of the most extreme SNR identified in NGC 4030. The bottom spectrum zooms in on the $y$-axis to facilitate the visualization of weaker features. All lines marked in the bottom panel are from [Fe{\,\sc ii}]. {\it Right:} Cut-outs of Fig.\ \ref{fig:SrcDetector} around the source.
 }
\label{fig:TheBeast}
\end{figure*}

Several other UFLOs in our list exhibit the features at $\sim 7300$ \AA\ seen in the examples shown in Fig.\ \ref{fig:ExampleSpectralExtraction}, some more clearly, others weaker. The most spectacular example is shown in Fig.\ \ref{fig:TheBeast}. The top panel shows its spectrum in an autoscale, while the bottom one expands the $y$-axis to allow a clearer view of the weaker features. 

Its long collection of emission lines includes numerous [Fe{\,\sc ii}] transitions (from 5158 to 9033 \AA), signaling the existence of an extended partially ionized zone typical of shocks, but not  of \hii regions, where the transition from ionized to neutral phases is sharp \citep{Osterbrock&Ferland06}. The anomalous strength of the \oi6300, 6364 doublet conveys the same message. Transitions from [Ni{\,\sc ii}], [Cr{\,\sc ii}], and [Cl{\,\sc ii}] are also detected. These lines are often detected in Galactic and extra-galactic SNR studies \citep[e.g.][]{Russel&Dopita90,Levenson+95,Fesen&Hurford96,Dopita+19}.

The region around 7300 \AA\ is of particular interest in this paper, as its contains strong features that we have seen before in the examples in Fig.\ \ref{fig:ExampleSpectralExtraction}. The three peaks seen in Fig.\ \ref{fig:TheBeast} in this range contain more than three lines. The blue peak at 7291 \AA\ is due to [Ca{\,\sc ii}], though its blue wing seems distorted by a bit of He{\,\sc i} 7281. The red one at $\sim 7381\,$\AA\ is likely due to a blend of [Ni{\,\sc ii}] 7378 and [Fe{\,\sc ii}] 7388. The central and strongest peak at $\sim 7323$ \AA\ blends a couple of \oii doublets (7319, 7320, and 7330, 7331) and possibly [Ca{\,\sc ii}] 7324. 
Disentangling these blended lines would require a detailed modelling \citep[say, as in][]{Dopita+19}.
Instead, we shall use the mere visual detection of these features as a signpost for SNRs and check how many more of our sources exhibit them.

The RGB composites in the right-hand panels of Fig.\ \ref{fig:TheBeast} show that this SNR lives in a region of ongoing star formation (at the tip of a spiral arm), suggesting that it had a massive progenitor star. 
These same images show another UFLO candidate at $(x, y) = (289, 112)$,  just $\sim 13$ spaxels (2.6 arcsec) down from the main source. This weaker source is not among the 26 in our primary sample, but it is picked up when the detection threshold is lowered to $T = 100 L_\odot$. Though its spectrum at $\sim 7300$ \AA\ is too noisy to make any conclusive statement about the presence or otherwise of SNR features, this source does have line ratios typical of SNRs. Indeed, it is included in the SNR sample introduced in the next section.


\section{Supernova Remnants in NGC 4030}
\label{sec:Results}

The analysis in the previous section provided strong evidence that many of our UFLOs are associated with SNRs. Some, however, are not, and hence our purely PCA-based detection criterion does not produce a clean list of SNRs. This is hardly surprising, as the method was not designed to achieve this specific goal. It is, nevertheless, easy to adapt our analysis to focus on SNRs. 

This section starts by examining whether  SNR candidates defined in terms of a canonical $\sii/\Ha$-based criterion are picked by our PCA-based method (Section \ref{sec:S2Detections}). After this basic test, we produce a list of SNR candidates which are detected through PC2  and satisfy a combined \sii/\Ha and \nii/\Ha criterion (Section \ref{sec:MixedCriterion}). The mean spectrum of the sources are presented in Section \ref{sec:SNRspectra}, where we also examine more closely the 7300 \AA\ features seen in the SNRs in Figs.\ \ref{fig:ExampleSpectralExtraction} and \ref{fig:TheBeast}. The density-sensitive $\sii 6716 / \sii 6731$ ratio and line widths of our SNRs are briefly discussed in  Section \ref{sec:S2ratio_and_vd}. Finally, Section \ref{sec:TheWeirdo} is dedicated to a peculiar source that is definitely not an SNR, but a very  interesting source nonetheless.

\subsection{[SII]-based search for SNRs}
\label{sec:S2Detections}

SNRs are traditionally identified  by their enhanced \sii lines, with \sii/\Ha ratios in excess of 0.4 \citep[e.g.,][]{Long+19}. To search for SNRs using this criterion we have first searched for unresolved sources in a  \sii map. In order to facilitate the identification of peaks we first subtract from the \sii image a smoothed version of itself and set negative entries to zero. As in Section \ref{sec:Detection}, DAOStarFinder was used to identify potential sources. Sources were then  extracted and only those with $\sii/\Ha \ge 0.4$ were retained. Candidates with very noisy spectra were manually removed from the list.
 
This process yielded 20 SNR-like sources. Only 8 of these are amongst the 26 UFLOs in our primary sample, but others are picked for lower detection thresholds. All but one are included in the $T = 100 L_\odot$ sample and the remaining source is picked lowering $T$ to $50 L_\odot$. To conclude, this exercise shows that our PCA-based definition of UFLOs can detect SNRs defined in a traditional way as long as the PC2 detection threshold is adjusted to $T = 50$--$100 L_\odot$.

\subsection{A refined list of SNR candidates}
\label{sec:MixedCriterion}

The flip side of PCA as a technique to find SNRs is that, as already pointed out, it picks many non-SNR sources as well. In this section we combine our PCA methodology with conventional emission-line ratio criteria to produce a list of SNR candidates. 

We keep tomogram 2 as a source detection image because of its much higher contrast in comparison to the \sii images, as illustrated in Fig.\ \ref{fig:BckExamples}. Moreover, tomogram 2 is not based on any single emission-line, but on all of them simultaneously. Correlations such as those between \nii and \sii emission (Fig.\ \ref{fig:BPT4UFLOs}) are thus automatically built in PC2, making it a robust image for source detection purposes.

Guided by the results of the previous section, let us now search for sources with SNR-like line ratios in the $T = 100 L_\odot$ UFLO sample. Only  $\sim 1/5$ of these UFLOs satisfy $\sii/\Ha > 0.4$, but this is likely a too stringent threshold. Many other sources have clearly enhanced \sii/\Ha and yet do not make this cut. In order to select SNR candidates we instead use the new \citet{Kopsacheili+20} classification scheme based on the combination of the \sii/\Ha and \nii/\Ha ratios (the dashed black curve in the bottom panel of Fig.\ \ref{fig:BPT4UFLOs}). This more inclusive criterion produces a list of  59 SNR candidates. Lowering the PC2 detection threshold to $T = 50 L_\odot$  adds 40 other sources to this list.

\begin{figure}
  \includegraphics[width=\columnwidth]{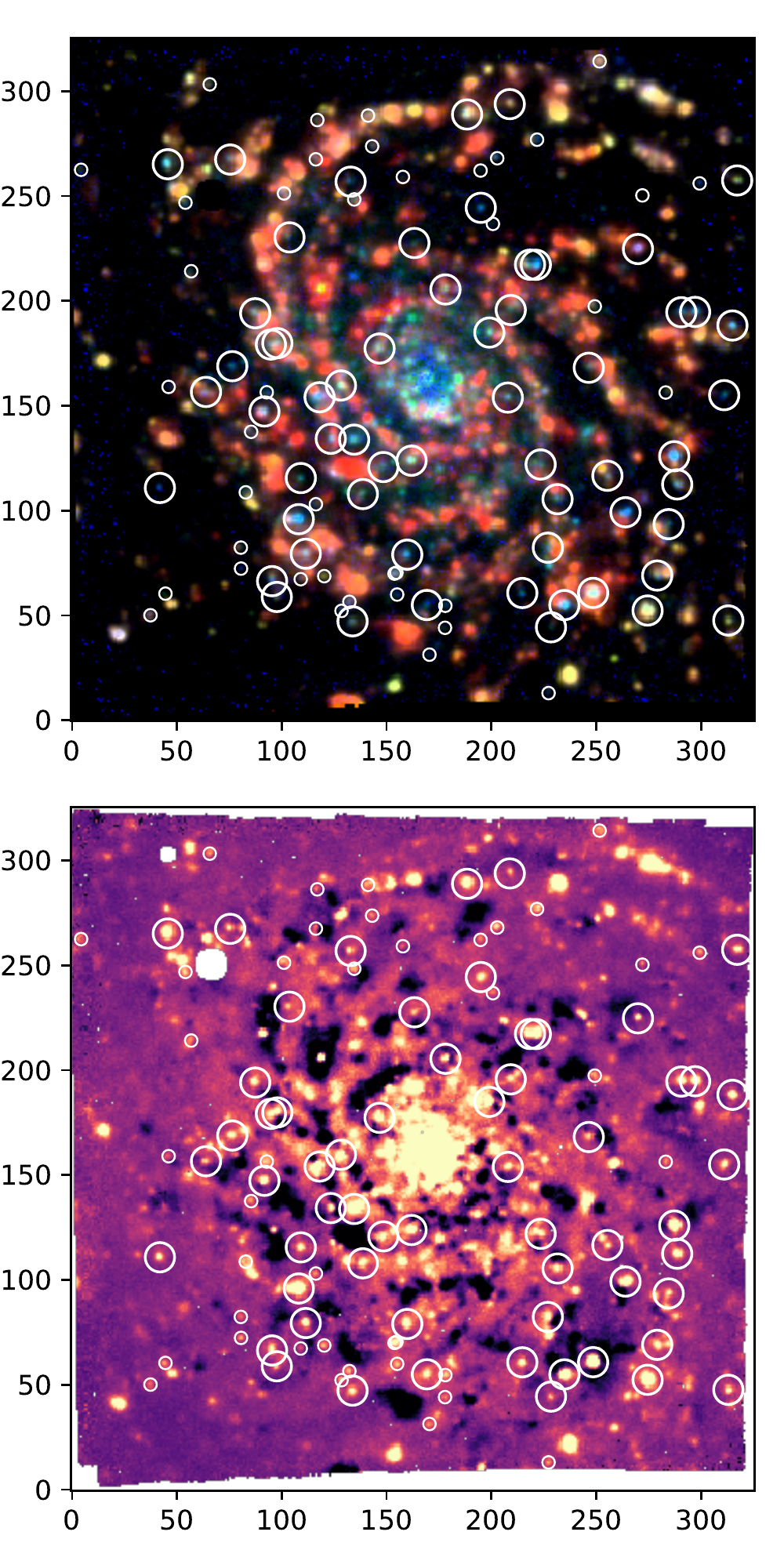}
  \caption{SNR sample sources plotted on the emission lines RGB composite (top, same as in Fig.\ \ref{fig:SrcDetector}) and  tomogram 2 (bottom, as in Fig.\ \ref{fig:Tomograms}). Large circles mark the 59 SNRs detected with a $T = 100 L_\odot$ threshold on tomogram 2. The 40 small circles are the ones further detected lowering $T$ to $50 L_\odot$.
}
\label{fig:KopSNRs_maps}
\end{figure}

\begin{figure*}
  \includegraphics[width=2\columnwidth]{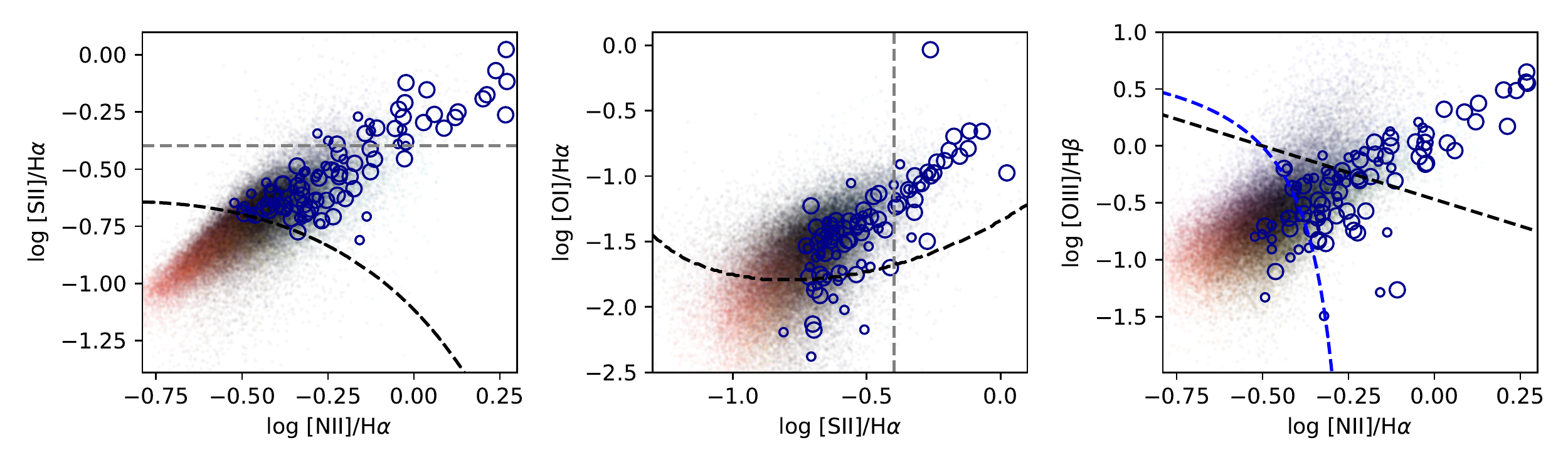}
  \caption{Diagnostic diagrams for the SNR sample.  Large circles mark the 59 SNRs detected with a $T = 100 L_\odot$ threshold. Small circles mark the 40 extra sources detected with $T = 50 L_\odot$. 
Black dashed lines in all diagrams show the \citet{Kopsacheili+20} SNR/\hii region demarkation lines. Other lines as in Figs.\ \ref{fig:BPT4UFLOs} and \ref{fig:[OI]}.
} 
\label{fig:KopSNRs_DDs}
\end{figure*}

Fig.\ \ref{fig:KopSNRs_maps} marks these sources in our emission-line RGB composite (top) as well as on tomogram 2 (bottom). Large circles correspond to the 59 sources found with $T = 100 L_\odot$, while small circles mark those found only in the $T = 50 L_\odot$ sample. Note that a few sources are hardly visible in the RGB frame, though all are clearly seen in tomogram 2. 

Fig.\ \ref{fig:KopSNRs_DDs} shows the loci of our sources in three diagnostic diagrams. By construction, all sources are above the \citet{Kopsacheili+20} SNR/\hii region division line in the \sii/\Ha vs.\  \nii/\Ha plane (left-hand panel). Most are also located above their division line in the \oi/\Ha vs.\  \sii/\Ha diagram (middle). 
Unsurprisingly, the top-right regions of these diagrams are populated exclusively by the subset of SNRs found with $T = 100 L_\odot$. Sources identified with a less stringent threshold in PC2 tend to have smaller forbidden-to-recombination line intensity ratios. 
For instance, the mean value of $\log\ \nii/\Ha$ for the $T = 100 L_\odot$ subsample is $-0.16$, while excluding these from the $T = 50 L_\odot$ list this average drops to $-0.29$. As expected, these subsets also differ in absolute terms. Line luminosities are on average about twice as bright for the $T = 100 L_\odot$ sources than for those only present in the $T = 50 \, L_\odot$ list.

The right-hand panel in Fig.\ \ref{fig:KopSNRs_DDs} shows our sources in the BPT diagram. Unlike in the \oi/\Ha vs.\  \sii/\Ha plane, the \citet{Kopsacheili+20} in the \oiii/\Hb vs.\  \nii/\Ha space does not separate well our objects, with only 27  out of the 59 sources in the $T = 100 L_\odot$ sample lying  in the SNR domain. As explained by \citet{Kopsacheili+20} themselves, this apparent inconsistency is to some extent expected, given the partial overlap in emission-line properties of \hii regions and SNRs. Still, this result raises an important question: {\em which diagnostic diagram should be used to pick SNR candidates?}

Observationally, the \sii/\Ha vs.\  \nii/\Ha diagram is undoubtedly the best choice for this particular study, as it comprises the strongest emission lines in NGC 4030. Theoretically, however, this is not an optimal choice.  Of the diagrams in  Fig.\ \ref{fig:KopSNRs_DDs} the  \oi/\Ha vs.\ \sii/\Ha one offers the best combination of completeness and contamination in separating SNRs from \hii regions according to \citet{Kopsacheili+20}, but the weakness of \oi is an obvious disadvantage in practical work. Similarly, the BPT diagram has a better performance in terms of contamination, but  the \oiii flux is generally low in NGC 4030 ($\sim 3$ times weaker than \sii in the median), and hence prone to uncertainties.

These considerations (1)  justify our choice of \sii/\Ha vs.\  \nii/\Ha diagram as a means to identify SNRs in NGC 4030, and (2) explain why we have not applied the theoretically more powerful 3D diagnostic criteria proposed by  \citet{Kopsacheili+20}. 
They also serve as a reminder that our list of SNRs may well be contaminated at some level due both to inherent diagnostic ambiguities and the aforementioned source extraction issues (Section \ref{sec:Background}). Future work will be required to confirm the less obvious candidates.

With these caveats in mind, let us henceforth focus on the 59 sources detected with $T = 100 L_\odot$, hereafter the ``SNR sample". Table \ref{tab:KopSample} lists some information on these sources.

\begin{table}
\begin{center}
\begin{tabular}{@{}lcccccrr@{}}
\hline\hline
ID & RA & Dec. & $\frac{\nii}{\Ha}$ & $\frac{\sii}{\Ha}$ & $\frac{\oi}{\Ha}$ & $\frac{\oiii}{\Hb}$ & $\log \frac{L_{\Ha}}{L_\odot}$ \\
\hline
1 & 22.90s & $6^\prime 23.8^{\prime\prime}$ & 0.93 & 0.53 & 0.11 & 0.94 & 3.23 \\
2 & 24.16s & $6^\prime 23.2^{\prime\prime}$ & 0.60 & 0.41 & 0.06 & 0.17 & 3.61 \\
3 & 21.77s & $6^\prime 23.2^{\prime\prime}$ & 0.46 & 0.17 & 0.00 & 0.15 & 3.93 \\
4 & 22.28s & $6^\prime 22.2^{\prime\prime}$ & 0.44 & 0.19 & 0.02 & 0.18 & 4.63 \\
5 & 23.69s & $6^\prime 21.7^{\prime\prime}$ & 1.59 & 0.64 & 0.16 & 3.10 & 3.09 \\
6 & 22.81s & $6^\prime 21.7^{\prime\prime}$ & 0.90 & 0.58 & 0.13 & 0.80 & 4.13 \\
7 & 24.64s & $6^\prime 20.9^{\prime\prime}$ & 0.46 & 0.23 & 0.04 & 0.15 & 3.72 \\
8 & 23.08s & $6^\prime 20.5^{\prime\prime}$ & 1.86 & 1.06 & 0.11 & 4.46 & 3.14 \\
9 & 22.63s & $6^\prime 20.5^{\prime\prime}$ & 0.65 & 0.29 & 0.04 & 0.54 & 4.54 \\
10 & 24.67s & $6^\prime 19.4^{\prime\prime}$ & 0.67 & 0.26 & 0.04 & 1.08 & 3.71 \\
11 & 22.22s & $6^\prime 18.9^{\prime\prime}$ & 0.38 & 0.22 & 0.02 & 0.23 & 4.14 \\
12 & 23.81s & $6^\prime 16.9^{\prime\prime}$ & 0.95 & 0.42 & 0.06 & 1.27 & 3.66 \\
13 & 24.46s & $6^\prime 16.8^{\prime\prime}$ & 0.67 & 0.34 & 0.07 & 0.85 & 3.82 \\
14 & 22.92s & $6^\prime 16.2^{\prime\prime}$ & 0.42 & 0.27 & 0.03 & 0.25 & 3.91 \\
15 & 22.15s & $6^\prime 14.0^{\prime\prime}$ & 0.60 & 0.24 & 0.05 & 0.53 & 3.66 \\
16 & 24.50s & $6^\prime 13.5^{\prime\prime}$ & 1.86 & 0.77 & 0.22 & 3.55 & 3.81 \\
17 & 22.42s & $6^\prime 12.8^{\prime\prime}$ & 1.07 & 0.51 & 0.09 & 2.10 & 3.77 \\
18 & 22.86s & $6^\prime 11.6^{\prime\prime}$ & 0.50 & 0.21 & 0.01 & 0.53 & 3.75 \\
19 & 24.09s & $6^\prime 11.1^{\prime\prime}$ & 0.78 & 0.48 & 0.07 & 0.05 & 3.54 \\
20 & 25.38s & $6^\prime 10.5^{\prime\prime}$ & 1.73 & 0.85 & 0.22 & 3.07 & 3.12 \\
21 & 22.09s & $6^\prime 10.2^{\prime\prime}$ & 0.61 & 0.37 & 0.04 & 0.50 & 3.87 \\
22 & 24.49s & $6^\prime 9.6^{\prime\prime}$ & 1.15 & 0.55 & 0.10 & 0.91 & 3.40 \\
23 & 22.54s & $6^\prime 9.4^{\prime\prime}$ & 0.46 & 0.27 & 0.03 & 0.33 & 3.84 \\
24 & 23.96s & $6^\prime 8.5^{\prime\prime}$ & 0.41 & 0.27 & 0.05 & 0.30 & 3.60 \\
25 & 22.96s & $6^\prime 8.3^{\prime\prime}$ & 0.75 & 0.39 & 0.02 & 1.18 & 3.40 \\
26 & 23.78s & $6^\prime 7.9^{\prime\prime}$ & 0.77 & 0.35 & 0.05 & 0.49 & 3.96 \\
27 & 22.11s & $6^\prime 7.5^{\prime\prime}$ & 1.85 & 0.55 & 0.92 & 3.64 & 3.95 \\
28 & 24.30s & $6^\prime 5.8^{\prime\prime}$ & 0.54 & 0.19 & 0.03 & 0.61 & 3.85 \\
29 & 24.15s & $6^\prime 5.9^{\prime\prime}$ & 1.32 & 0.53 & 0.03 & 1.63 & 3.44 \\
30 & 24.72s & $6^\prime 3.2^{\prime\prime}$ & 0.42 & 0.22 & 0.03 & 0.45 & 4.15 \\
31 & 23.17s & $6^\prime 1.9^{\prime\prime}$ & 0.94 & 0.35 & 0.07 & 1.08 & 3.48 \\
32 & 24.37s & $6^\prime 1.9^{\prime\prime}$ & 0.61 & 0.30 & 0.04 & 0.75 & 3.96 \\
33 & 21.80s & $6^\prime 1.7^{\prime\prime}$ & 1.63 & 0.67 & 0.20 & 1.49 & 3.22 \\
34 & 25.09s & $6^\prime 1.4^{\prime\prime}$ & 0.39 & 0.25 & 0.02 & 0.42 & 3.88 \\
35 & 24.23s & $6^\prime 0.7^{\prime\prime}$ & 0.63 & 0.24 & 0.03 & 0.27 & 4.02 \\
36 & 22.66s & $5^\prime 59.1^{\prime\prime}$ & 0.38 & 0.21 & 0.03 & 0.19 & 3.94 \\
37 & 24.92s & $5^\prime 58.9^{\prime\prime}$ & 0.88 & 0.48 & 0.05 & 1.09 & 3.35 \\
38 & 23.99s & $5^\prime 57.2^{\prime\prime}$ & 0.52 & 0.23 & 0.04 & 0.24 & 3.91 \\
39 & 24.68s & $5^\prime 56.8^{\prime\prime}$ & 0.49 & 0.21 & 0.04 & 0.45 & 3.83 \\
40 & 24.64s & $5^\prime 56.7^{\prime\prime}$ & 0.53 & 0.29 & 0.02 & 0.39 & 3.79 \\
41 & 23.29s & $5^\prime 55.7^{\prime\prime}$ & 0.56 & 0.23 & 0.03 & 0.27 & 3.90 \\
42 & 21.74s & $5^\prime 55.0^{\prime\prime}$ & 0.95 & 0.76 & 0.16 & 0.71 & 3.78 \\
43 & 24.78s & $5^\prime 53.9^{\prime\prime}$ & 0.47 & 0.26 & 0.03 & 0.31 & 3.95 \\
44 & 22.07s & $5^\prime 53.8^{\prime\prime}$ & 0.72 & 0.45 & 0.08 & 0.82 & 3.57 \\
45 & 21.98s & $5^\prime 53.7^{\prime\prime}$ & 1.09 & 0.70 & 0.14 & 1.06 & 3.64 \\
46 & 23.15s & $5^\prime 53.6^{\prime\prime}$ & 0.57 & 0.31 & 0.04 & 0.21 & 3.61 \\
47 & 23.57s & $5^\prime 51.6^{\prime\prime}$ & 0.58 & 0.20 & 0.06 & 0.18 & 3.79 \\
48 & 23.03s & $5^\prime 49.2^{\prime\prime}$ & 0.61 & 0.29 & 0.04 & 0.51 & 3.87 \\
49 & 22.99s & $5^\prime 49.2^{\prime\prime}$ & 1.22 & 0.48 & 0.10 & 1.99 & 3.77 \\
50 & 22.34s & $5^\prime 47.7^{\prime\prime}$ & 0.37 & 0.21 & 0.02 & 0.63 & 4.18 \\
51 & 23.77s & $5^\prime 47.2^{\prime\prime}$ & 0.48 & 0.22 & 0.03 & 0.20 & 3.49 \\
52 & 24.56s & $5^\prime 46.6^{\prime\prime}$ & 0.32 & 0.20 & 0.01 & 0.20 & 3.86 \\
53 & 23.34s & $5^\prime 43.8^{\prime\prime}$ & 1.34 & 0.56 & 0.11 & 2.37 & 3.31 \\
54 & 24.17s & $5^\prime 41.3^{\prime\prime}$ & 0.75 & 0.31 & 0.06 & 1.00 & 3.43 \\
55 & 21.71s & $5^\prime 41.2^{\prime\prime}$ & 0.48 & 0.20 & 0.01 & 0.14 & 3.96 \\
56 & 25.33s & $5^\prime 39.6^{\prime\prime}$ & 0.94 & 0.62 & 0.13 & 0.69 & 3.91 \\
57 & 24.94s & $5^\prime 39.2^{\prime\prime}$ & 0.46 & 0.33 & 0.05 & 0.23 & 3.83 \\
58 & 23.43s & $5^\prime 34.9^{\prime\prime}$ & 0.47 & 0.24 & 0.04 & 0.26 & 4.25 \\
59 & 23.16s & $5^\prime 33.9^{\prime\prime}$ & 0.34 & 0.20 & 0.01 & 0.08 & 4.05 \\
\hline
\end{tabular}
\end{center}
\caption{
    Coordinates and emission-line properties for sources in the SNR sample. 
    Coordinates are given as offsets from RA $=$ 12$^{\rm h}$00$^{\rm m}$00$^{\rm s}$ and  Dec.\ $= -01^\circ 00^\prime 00^{\prime\prime}$.
}
\label{tab:KopSample}
\end{table}

\subsection{Spectra and the 7300 \AA\ features}
\label{sec:SNRspectra}

\begin{figure*}
  \includegraphics[width=2\columnwidth]{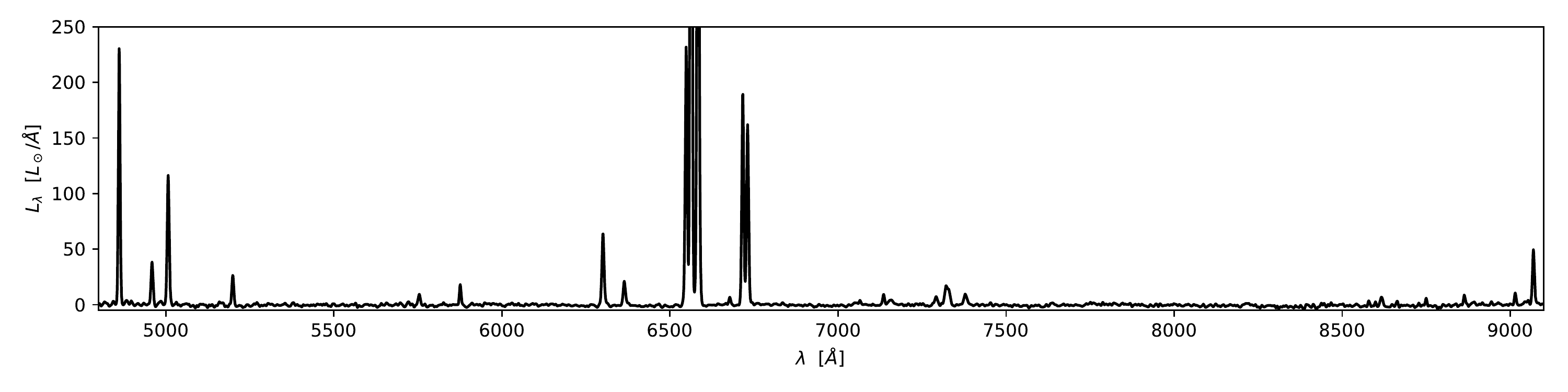}
  \caption{Mean spectrum of the 59 sources in the SNR sample. 
} 
\label{fig:KopSNRs_MeanSpectrum}
\end{figure*}

Fig.\ \ref{fig:KopSNRs_MeanSpectrum} shows the mean spectra of the 59 sources in the SNR sample. The meaning of such an average is of course not clear, since it mixes not only SNRs with different excitation conditions but also differently reddened by dust. None the less the increased signal-to-noise is useful to pick weak emission lines that are not obvious in the individual spectra.

Zooming on the $y$-scale one can identify most of the lines discussed in connection with Fig.\ \ref{fig:TheBeast}, where our most extreme SNR was first presented, albeit with smaller intensity. We note in passing that that source, as well as the two other examples shown in Fig.\ \ref{fig:ExampleSpectralExtraction}, all satisfy the emission-line based criteria applied to clean our PC2-based sample of non-SNR sources.

\begin{figure}
  \includegraphics[width=\columnwidth]{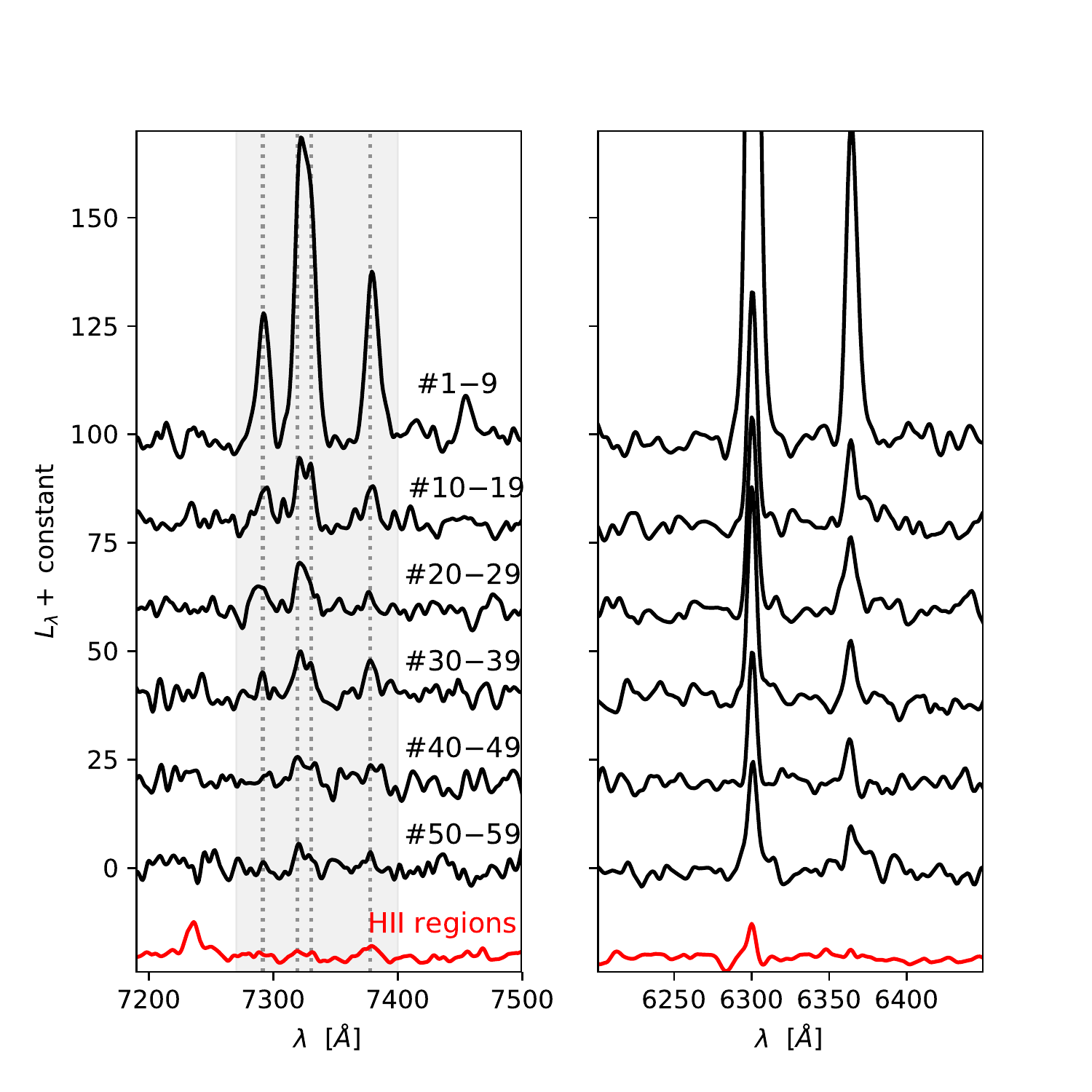}
  \caption{Average spectra obtained by dividing our sorted list of 59 sources into six bins. The left-hand panel focus on the 7330 \AA\ features. Numbers towards the right indicate which of the sources in the sorted list are included in each bin.
The right-hand panel zooms into the \oi$\lambda\lambda6300,6363$ lines for comparison. In both cases the bottom spectrum (in red) represents an average over \hii regions in NGC 4030.
 }
\label{fig:KopSNRs_7300stacks}
\end{figure}

The features at $\sim 7300$ \AA\ (including lines from [Ca{\,\sc ii}], \oii, [Ni{\,\sc ii}], and [Fe{\,\sc ii}]; see Section \ref{sec:TheBeast}) are clearly present in the mean spectrum. The fact that these features have been previously observed in studies of Galactic and extragalactic SNRs further strengthens the identification of our sources as SNRs. Yet, these lines are only clearly visible in about 10 of our sources, the most spectacular example being the case shown in Fig.\ \ref{fig:TheBeast}. For the remaining ones they are either not present or too weak and immersed in the noise.

To investigate this further we sorted the 59 sources by an index that quantifies the excess emission in the 7270--7400 \AA\ range and stacked the spectra into six groups, each containing 9 or 10 sources. Fig.\ \ref{fig:KopSNRs_7300stacks} (left-hand panel) shows the result of this experiment. The top spectrum is the average of the nine sources with the strongest 7300 \AA\  features, the second one is the average of numbers 10 to 19 in the sorted list, and so on. By construction, the lines become weaker as one moves towards the bottom of the plot. The increased signal-to-noise of these stacks allows us to detect the lines down to the fourth bin, after which there are only inconclusive hints of their presence.
For comparison, the right-hand panel in Fig.\ \ref{fig:KopSNRs_7300stacks} shows the same sequence of stacked spectra but around the  \oi$\lambda\lambda6300,6363$ doublet, and using the same luminosity density scale as in the left-hand panel. 
This comparison shows that the strength of the lines around 7300 \AA\ decreases in tandem with the decrease of  the \oi lines (and other lines in general). We speculate that this progression towards weaker SNR features is due to evolution of the remnants as they gradually exhaust their finite energy content.

What do \hii regions in NGC 4030 look like in this spectral window?
In order to answer this question we have culled a sample of \hii regions from the MUSE data and evaluated their stacked spectra. The 143 \hii regions used for this test were identified with a simple peak finding algorithm applied to an \Ha image where spaxels with non-\hii region like line ratios were excluded -- a maximum value of 0.4 was admitted for the \nii/\Ha, \oiii/\Hb, and \sii/\Ha. Regions around all $T = 100 L_\odot$ sources were also masked to avoid contamination, but otherwise the spectra were extracted as for other sources in this paper.
The bottom (red) spectrum in Fig.\ \ref{fig:KopSNRs_7300stacks} shows the result of this analysis. The plot shows that, at least in NGC 4030, \hii regions have negligible emission features around 7300 \AA. \citet{Levenson+95}, in their study of the N63A nebula in the Large Magellanic Cloud, also find that its \hii region component has much weaker 7300 \AA\ features than the SNR within the system.

To summarize, the 7300 \AA\ features are individually detected in $\sim$ 10 sources, and statistically in $\sim 40$. The comparison with the progression of the \oi lines in Fig.\ \ref{fig:KopSNRs_7300stacks} suggests that these lines may well be present even for the remaining sources, but at levels that do not allow us to detect them even statistically.

\subsection{[S{\small II}] ratio and line width}
\label{sec:S2ratio_and_vd}

\begin{figure}
  \includegraphics[width=\columnwidth]{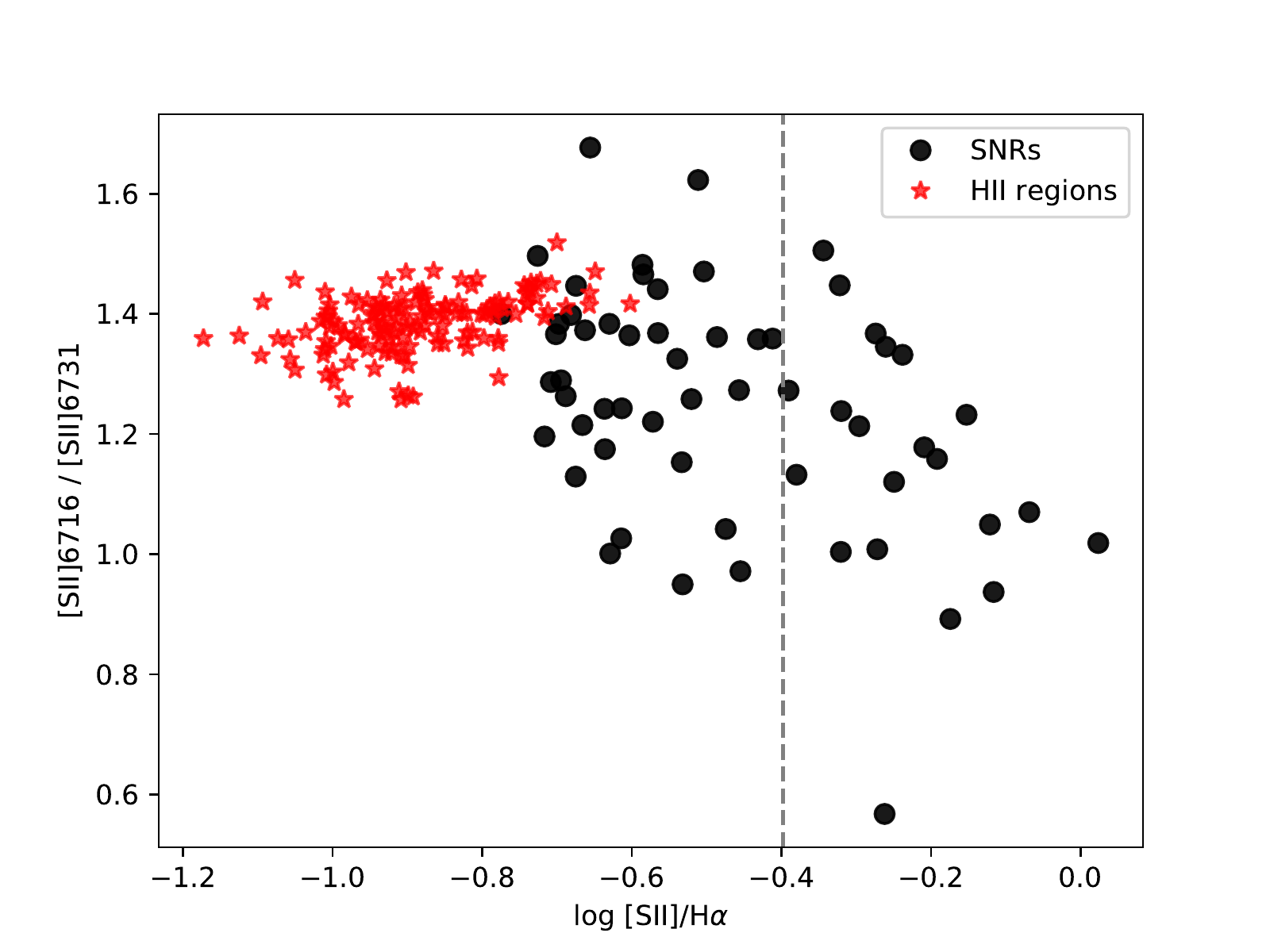}
  \caption{
 The density-sensitive \sii 6716/6731 ratio plotted against \sii/\Ha for SNRs and  \hii regions in NGC 4030. 
   }
\label{fig:S2ratio}
\end{figure}

Before closing, let us mention two other indications that the sources we have identified in this study are indeed SNRs. The first one comes from the density sensitive $\sii 6716 / \sii 6731$ ratio. At least for young remnants, one expects to find larger densities (smaller $\sii 6716 / \sii 6731$) than in \hii regions \citep[e.g.][]{Sabbadin+77,Leonidaki+13,Moumen+19}.
The \hii regions used to test  this are the same ones defined in Section \ref{sec:SNRspectra}, 
except that no background subtraction was applied in this case. 
Results of this comparison are shown in Fig.\ \ref{fig:S2ratio}. The plot shows that, despite some overlap, our SNRs clearly tend to have denser nebulae than \hii regions.\footnote{We note in passing that the sixth principal component expresses an anticorrelation between the \sii 6716 and 6731 lines, such that its image could, in principle, be used to trace dense spots. The peaks seen in tomogram 6 indeed coincide with our most extreme SNRs, but the image as a whole is very noisy given that it corresponds to the last PC.} In particular, the source with the lowest \sii ratio is the SNR in Fig.\ \ref{fig:TheBeast}. 

The second indication comes from line widths. The most obvious SNRs in our sample have emission lines that are visibly broader than those around them, but for many the difference is not so evident. Statistically, however, the difference is there. In nearly all (55/59) cases we find the mean width of the \nii6584 (the strongest forbidden line) within the source aperture to be larger than that on the corresponding background annulus, with an average difference of $26$ km$\,$s$^{-1}$ in terms of FWHM. One should however note that in approximately half of the cases the lines are not actually resolved at the instrumental resolution of 140 km s$^{-1}$ at the wavelength of \nii in MUSE. In any case, this difference is qualitatively consistent with the expectation that gas velocities are larger in SNRs than in their immediate surroundings.

\subsection{Serendipitous discovery of an SN impostor?}
\label{sec:TheWeirdo}

\begin{figure*}
  \includegraphics[width=2\columnwidth]{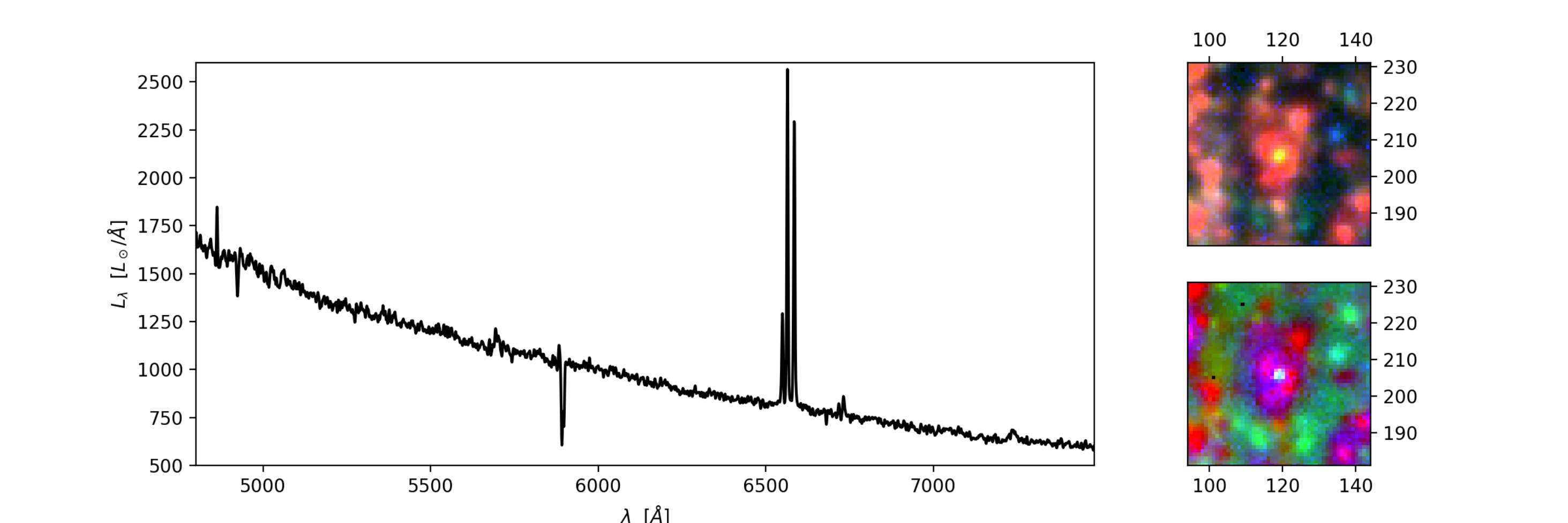}
  \caption{Spectrum of ``the beast", a serendipitously discovered $M_V \sim -12.5$ luminous blue variable on top of an \hii region in NGC 4030. Images on the right-hand panels are cut-outs of the RGB composites of Figs.\ \ref{fig:RGBlines} and \ref{fig:Tomograms}. }
\label{fig:TheWeirdo}
\end{figure*}

In the process of screening our PCA-based list of UFLOs for sources with SNR-like line ratios a very peculiar object was dropped from the sample. Signalled as an outlier in Fig.\ \ref{fig:BPT4UFLOs} because of its weak \oiii and \sii despite strong \nii emission, this source does not show \oi nor the 7300 \AA\ features. Located at pixel $(x,y) = 119, 205$ in Fig.\ \ref{fig:RGBlines}
($\mathrm{RA} = 12^\mathrm{h}00^\mathrm{m}24.36^\mathrm{s}$,  $\mathrm{Dec.} = -1^\circ05{}^\prime51.4{}^{\prime\prime}$), ``the beast", as we have nicknamed it,  is also peculiar in a fundamental aspect: Unlike other UFLOs and SNRs, it has a strong continuum.

Fig.\ \ref{fig:TheWeirdo} shows its spectrum, as extracted directly from the observed cube. As in Section \ref{sec:Background}, the spectral extraction was performed on a circle  $r_{\rm src} = 2.55$ spaxels wide, and the background was that estimated from the annulus between $r = 3$ and 6 spaxels. Note that the beast is located right on top of an \hii region, as can be seen from the cut-outs of the emission-line and PC RGB composites shown in Fig.\ \ref{fig:TheWeirdo}. 

The beast exhibits a very blue continuum, strong \nii and \Ha, weak \Hb and \sii emission. Other noticeable features include (1) a hint of a faint broad component in \Ha, (2) a strong Na{\,\sc i} D doublet absorption (3.5 \AA\ in equivalent width), (3) HeI 4921 and 6678 in absorption and 5876 in emission, (4) weak broad emission features at $\sim 5696$ and 7236 \AA\ that could be associated with C{\,\sc iii} and C{\,\sc ii}, respectively. {\em All in all it seems that we are looking not at an SNR, but at   a very luminous star}.

From the spectrum in Fig.\ \ref{fig:TheWeirdo} we estimate an astounding V band absolute magnitude of $-12.5$. Blackbody fits to its spectrum indicate an effective temperature of $\sim 10000$ K and an inferred radius of the order of $600 R_\odot$.
We have also verified that the source is variable. The datacube analysed throughout this paper is a combination of observations taken on Feb/6 and Mar/10 of 2016. Comparing these cubes we find that the beast brightened by $\sim 4.4$ percent at 5200 \AA\ in the 34 days between the two observations. This may not seem much, but  (1) it is the largest variation found in the whole FoV in both absolute and percentual terms, and (2) it is a $3.7 \sigma$ variation, in the sense that the beast brightened by 3.7 times the rms variation of all 26 UFLOs in our primary sample, none of which varies by more than $2\sigma$. 

The beast is thus luminous blue, and variable. Because of its extreme luminosity, we speculate that it is not just an LBV star, but an SN impostor \citep{VanDyk+2000,Smith+11}. \citet{Thone+17}
recently presented an extensive study on one such case: SN2015gb, an extragalactic LBV that exhibited extreme luminosity and variability for at least 20 yr before undergoing what seemed to be a terminal explosion as a type IIn SN in 2015.

Clearly, this extraordinary source deserves a dedicated study on its own.


\section{Discussion}
\label{sec:Discussion}

Now that we have identified a few dozen SNRs in NGC 4030, let us open a parenthesis outside  the original scope of this paper to evaluate what impact these sources may have on studies of star-forming regions in galaxies.

\subsection{Impact on standard emission-line diagnostics}
\label{sec:Impact}

Much of our knowledge about star-forming galaxies comes from nebular diagnostics based on the hypothesis that all emission lines are powered by young stars. Other sources of line emission, like SNRs, act as contaminants which induce biases in the estimates of properties like the star formation rate (SFR) and nebular metallicity ($Z_{\rm neb}  \equiv 12 + \log {\rm O/H}$).

In order to assess to which extend SNRs affect the SFR estimates in NGC 4030 we have computed the total \Ha luminosity with and without SNRs. We find that only 0.7\% of the \Ha flux in the FoV comes from our 59 SNRs, which translates into a negligible bias in $L_{\Ha}$-based SFR estimates. 
This fraction is in the range of those derived by  \citet{Vucetic+15}, who have evaluated the contribution of SNRs to the \Ha flux in 18 galaxies, obtaining fractions ranging from 0.1 to 12.8\%.

We have repeated the with vs.\ without SNRs calculations for galaxy-wide 
${\rm N2} \equiv \log \{ \nii/\Ha \}$ and ${\rm O3N2} \equiv \log \{ (\oiii/\Hb)/(\nii/\Ha) \}$ 
indices, two popular strong-line calibrators for the nebular metallicity \citep[][and references therein]{Curti+17}. Predictably, the effects are smaller than the already small effect on \Ha, since now the corrections affect both numerator and denominator. Indeed, we find that N2 and O3N2 are only affected in their third significant digits.

SNRs thus have a small effect in global SFR estimates and a negligible one in metallicity. {\em Locally}, however, their effects can be significant. To illustrate this we have defined regions of 2.8 arcsec (14 spaxels, or 78 pc) in diameter centred on the sources in our SNR sample. The N2 and O3N2 indices were then computed for the region as a whole and excluding the SNRs.

On average, \nii/\Ha comes out 1.069 times higher when the SNRs are not masked, a bias that translates to $\Delta Z_{\rm neb}({\rm N2}) = 0.022$ dex using the $Z_{\rm neb}({\rm N2})$ calibration of \citet{Curti+17}. Considering only the top 10 sources in terms of \sii/\Ha the bias in metallicity increases to  $\Delta Z_{\rm neb}({\rm N2}) = 0.041$ dex, while for
the SNR in Fig.\ \ref{fig:TheBeast} alone it reaches 0.071 dex. Numerically, these biases are of the same order as $Z_{\rm neb}$ variations within galaxies.

O3N2-based metallicities are much less sensitive to contamination by SNRs. The reason is evident from the BPT diagram itself (Fig.\ \ref{fig:BPT4UFLOs}, top), where one sees that SNRs move away from \hii regions along lines of roughly constant O3N2. (That is the same reason why DIG-related $Z_{\rm neb}$ corrections are smaller for O3N2; \citealt{ValeAsari2019}). Accordingly, we obtain an insignificant $\Delta Z_{\rm neb}({\rm O3N2}) = -0.0015$ dex for the sample as a whole. Even in the extreme case of Fig.\ \ref{fig:TheBeast} the bias is just $-0.019$ dex.

\subsection{The nature of HII regions with anomalous forbidden lines}
\label{sec:AnomalousHIIregions}

The numerical experiments reported just above emulate a situation where an unnoticed SNR contaminates the emission lines from what was presumed to be an area composed solely of \hii regions. While the use of circular apertures is obviously a simplification of the problem, more advanced methods to delineate \hii regions spatially \citep[e.g.][]{Sanchez+12b,Casado+2017,RousseauN+2018} would probably also include SNRs within the resulting contour, specially if observed under coarser resolution.

The possibility that SNRs contaminate the line emission from \hii regions has long been acknowledged in the literature. \citet{Kennicutt+89}, for instance, discussed it very didactically over 30 yr ago in a comparison of nuclear and disc \hii regions. More recently, integral field spectroscopy work with CALIFA and MUSE identified numerous examples of emission regions well outside the nuclei of galaxies that look just like \hii regions except for somewhat enhanced forbidden lines, as seen in the studies by \citet{Sanchez+12a}, \citet{Sanchez+15}, and \citet{Sanchez-Menguiano+20}.

In this context it is appropriate to point out that we have explicitly demonstrated that, at least in NGC 4030, contamination by SNRs leads to such ``anomalous" \hii regions. As illustrated by virtually every image shown in this study (starting with Fig.\ \ref{fig:RGBlines}), SNRs are very often surrounded by or immersed within star-forming regions. Then, when examining their positions in diagnostic diagrams prior to background subtraction (red stars in Figs.\ \ref{fig:BPT4UFLOs} and \ref{fig:[OI]}), we saw that, besides the obvious non-\hii cases (the clearest SNR detections), some populate regions that slightly trespass the \hii region demarcation frontier due to the SNR component. Source 31 in Table \ref{tab:KopSample} is a good example. Its raw (uncorrected for background) BPT coordinates are ($\log \nii/\Ha, \log \oiii/\Hb) = (-0.34,-0.31)$, a bit above the \citet{Stasinska+06} limit. After background removal its line ratios move into the realm of SNRs, as indicated by the arrows in Fig.\ \ref{fig:BPT4UFLOs}. Conversely, removing the SNR component moves them to $(-0.50, -0.53)$, well within the star-forming zone.

The fact that we were able to isolate the SNR contribution in such complex star-forming environment is ultimately due to the superb spatial sampling of MUSE. Observed under the much coarser resolutions of CALIFA or MaNGA  most of the SNRs identified in this study would not be recognized as individual sources, leaving only trace indications of their existence in the form of slight ``anomalies" in diagnostic diagrams.


\section{Summary}
\label{sec:Conclusions}

We have reported the discovery of SNRs based on MUSE integral field spectroscopy of NGC 4030. As the storyline of this paper reveals, this discovery came not from an pre-planned search for SNRs, but from an exploratory investigation on the nature of compact sources with enhanced forbidden line emission (UFLOs) spotted in a cursory examination of  emission-line images derived from the data (Fig.\ \ref{fig:RGBlines}). Here is a summary of how this investigation unfolded.

\begin{enumerate}

\item
PCA tomography of the \Hb, \oiii5007, \Ha, \nii6584, \sii6716, and 6731 emission lines (the strongest in the MUSE range) proved to be a useful tool to locate these sources. The second eigen line spectrum obtained from this analysis contrasts \oiii, \nii, and \sii with  \Hb and \Ha (Fig.\ \ref{fig:eigSpec}), such that UFLOs stand out clearly in the image of this PC (Fig.\ \ref{fig:Tomograms}).

\item
Due to the intense nebular emission throughout the field of view, extracting the source properties 
proved far more complicated than spotting them (Fig.\ \ref{fig:BckExamples}). 
To mitigate this problem our aperture photometry removes spaxels with excessive \Ha contribution to the background level.

\item For about half of the UFLOs the background subtraction leads to an increase in standard diagnostic flux ratios of collisional to recombination lines (\nii/\Ha, \sii/\Ha, \oiii/\Hb, \oi/\Ha; Figs.\ \ref{fig:BPT4UFLOs} and \ref{fig:[OI]}), making them even more different from the \hii regions that dominate the nebular emission in NGC 4030.  Others, however, have \hii-region-like line ratios.

\item
The smoking gun evidence that at least some UFLOs are associated with SNRs came from the spectral extraction (Figs.\ \ref{fig:ExampleSpectralExtraction} and \ref{fig:TheBeast}), which
revealed  anomalous \nii, \sii and \oi emission as well as features around 7300 \AA\  (including lines from [Ca{\,\sc ii}], \oii, [Ni{\,\sc ii}], and [Fe{\,\sc ii}]) which distinguish them from \hii regions. These same features, all of which are indicative of shocks, have been previously identified in several studies of Galactic and extragalactic SNRs, but never before used as a diagnostic. 

\item 
We have screened our list of PCA-based detections for sources with SNR-like emission-line ratios. A total of 59 sources fall in the SNR region of the \sii/\Ha vs.\ \nii/\Ha diagram (Figs. \ref{fig:KopSNRs_maps} and  \ref{fig:KopSNRs_DDs}) according to the recently proposed  \hii region/SNR separation line by \citet{Kopsacheili+20}.

\item
The $\sim 7300$ \AA\ SNR features are evident in the mean spectrum (Fig.\ \ref{fig:KopSNRs_MeanSpectrum}) as well as in $\sim 10$ individual sources. They are also apparently present (though weaker) in other sources. A stacking analysis reveals that indeed these features are present in about 40 objects (Fig.\ \ref{fig:KopSNRs_7300stacks}). The remaining sources also have weaker emission lines in general, so the 7300 \AA\ lines may well be present but at undetectable flux levels.

\item The impact of SNRs on global SFR and nebular metallicity estimates in NGC 4030 is small. Locally, ignoring their presence may lead to significant biases in $Z_{\rm neb}$ estimates based on the N2 index, while O3N2-based metallicities are not affected.

\item 
Another local effect of SNRs is that, when superposed to \hii regions, they may add enough forbidden lines to move the source beyond the limits for star-forming regions devised on the basis of photoionization models. This situation (envisaged over three decades ago) happens in the NGC 4030 data examined here, and is a likely explanation for the so called ``anomalous" \hii regions found in recent integral field studies.

\item 
An unexpected result was the serendipitous discovery of a luminous blue variable star which seems to fit the profile of a SN impostor (Fig.\ \ref{fig:TheWeirdo}), and whose nature deserves further investigation.

\end{enumerate}

Our search for the nature of the intriguing green/blue compact (unresolved) sources in Fig.\ \ref{fig:RGBlines} has thus led us to the discovery of about five dozen SNRs in NGC 4030. The true number of SNRs may well be larger, of course, and in fact, some promising candidates can be found among the candidates detected with a threshold on PC2 lower than the one used for our SNR sample. 

The rather unorthodox methodology employed here has plenty of room for improvement, specially as it was not originally designed to find SNRs. With the benefit of hindsight, one can envisage ways of improving upon the PCA tomography technique as used in this paper. For instance, including weaker emission lines 
like \oi and the 7300 \AA\ features
in the analysis would clearly help identifying SNRs, though one would then have to deal with very uncertain or missing information, perhaps with techniques such as those described in \citet{Budavari+2009}.
Improvements on the source extraction strategy should also be attempted, as this is critical to the diagnostic of  physical properties of individual SNRs (e.g., shock velocities, densities). Combining such tools with the collection of SNRs reported here, presumably each with a different age, should provide useful empirical information to better understand the evolution of SNRs. 

Finally, it is worth noting that these are the most distant optically detected SNRs to date. According to \citet{Vucetic+15} the record belonged to NGC 2903, a galaxy 9 Mpc away. At 29 Mpc, NGC 4030 triples this distance. This record should not hold for long, however, as there are several other galaxies in the MUSE archive which can be dissected with the same kind of methodology employed in this paper.


\section*{Acknowledgements}

The authors would like to thank Christina Th\"one, Enrique P\'erez, Gra\.zyna Stasi\'nska, and Llu\'{\i}s Galbany for valuable discussions. 
We are in great debt to the (anonymous) referee for her/his very thorough report.
Likewise, we would like to thank the recently deceased Prof.\ Jo\~ao Steiner for the outstanding and inspiring work throughout his career.
RCF acknowledges support from CNPq. SFS thanks   CONACYT FC-2016-01-1916 and CB285080 projects and PAPIIT IN100519 project for support on this study.

\section*{DATA AVAILABILITY}

The data used in this work are available from the ESO Science Archive at https://archive.eso.org/.

\bibliography{NGC4030_references}

\end{document}